\documentclass[useAMS,usenatbib,usegraphicx]{mn2e}
\usepackage{util}
\usepackage{txfonts}
\usepackage{multirow}

\title[WR stars and the ionized ISM in the metal-poor WR galaxy Mrk~178]{Uncovering multiple Wolf-Rayet star-clusters and the ionized ISM in Mrk~178:
 the closest metal-poor Wolf-Rayet HII galaxy}
\author[C. Kehrig et al.]{C. Kehrig$^{1}$\thanks{E-mail:kehrig@iaa.es}, E. P\'erez-Montero$^{1}$,  J.M. V\'ilchez$^{1}$,  J. Brinchmann$^{2}$, D. Kunth$^{3}$,
\newauthor
R. Garc\'ia-Benito$^{1}$, P. A. Crowther$^{4}$, J. Hern\'andez-Fern\'andez$^{5}$,
F. Durret$^{3}$, T. Contini$^{6,7}$,
\newauthor
A. Fern\'andez-Mart\'in$^{1}$  and B.L. James$^{8}$\\
$^{1}$Instituto de Astrof\'isica de Andaluc\'ia, CSIC, Apartado de correos 3004, 18080 Granada, Spain \\
$^{2}$ Leiden Observatory, Leiden University, PO Box 9513, 2300 RA Leiden, The Netherlands \\
$^{3}$ Institut d'Astrophysique de Paris, UMR 7095 CNRS, Universit\'e Pierre \& Marie Curie, 98 bis boulevard Arago, 75014 Paris, France  \\ 
$^{4}$ Department of Physics \& Astronomy, University of Sheffield,
Hicks Building, Hounsfield Road, Sheffield S3 7RH, United Kingdom \\
$^{5}$ Departamento de Astronomia, Instituto de Astronomia,Geof\'isica e Ci\^encias Atmosf\'ericas da USP, Rua do Mat\~ao 1226, Cidade Universit\'aria 05508-090 S\~ao Paulo, Brazil \\
$^{6}$ Institut de Recherche en Astrophysique et Plan\'etologie (IRAP), CNRS, 14 avenue \'Edouard Belin, F-31400  Toulouse, France \\
$^{7}$ IRAP, Universit\'e de Toulouse, UPS-OMP, Toulouse, France \\
$^{8}$ Institute of Astronomy, University of Cambridge, Madingley Road, Cambridge, CB3 0HA, United Kingdom}
\begin{document}

\date{Accepted Date. Received Date; in original Date}

\pagerange{\pageref{firstpage}--\pageref{lastpage}} \pubyear{2013}

\maketitle

\label{firstpage}

\begin{abstract}
  New integral field spectroscopy (IFS) has been obtained for the nearby
  metal-poor Wolf-Rayet (WR) galaxy Mrk~178 to examine the spatial
  correlation between its WR stars and the neighbouring ionized
  interstellar medium (ISM). The strength of the broad WR features and
  its low metallicity make Mrk~178 an intriguing object. We have
  detected the blue and red WR bumps in different locations across the
  field-of-view ($\sim$ 300 pc $\times$ 230 pc) in Mrk~178.  The study
  of the WR content has been extended, for the first time, beyond its
  brightest star-forming knot uncovering new WR star-clusters.  Using
  SMC/LMC-template WR stars we empirically estimate a minimum of
  $\sim$ 20 WR stars within the region sampled.  Maps of the spatial
  distribution of the emission-lines and of the physical-chemical
  properties of the ionized ISM have been created and analyzed.  Here
  we refine the statistical methodology by P\'erez-Montero et
  al. (2011) to probe the presence of variations in the ISM
  properties. An error-weighted mean of 12+log(O/H)=7.72 $\pm$ 0.01 is
  taken as the representative oxygen abundance for Mrk~178. A
  localized N and He enrichment, spatially correlated with WR stars,
  is suggested by this analysis.  Nebular He{\sc ii}$\lambda$4686 emission is shown
  to be spatially extended reaching well beyond the location of the WR
  stars. This spatial offset between WRs and He{\sc ii} emission can
  be explained based on the mechanical energy input into the ISM by
  the WR star winds, and does not rule out WR stars as the He{\sc ii}
  ionisation source.  We study systematic aperture effects on the
  detection and measurement of the WR features, using SDSS spectra
  combined with the power of IFS. In this regard, the importance of
  targeting low metallicity nearby systems is discussed.
\end{abstract}
%
%
\section{Introduction}\label{intro} 

HII galaxies are local, gas-rich, systems
with observable properties dominated by young and
massive star clusters \citep[e.g.,][]{M85,campos93,popescu00,K04,W04}.
HII galaxies are also low metallicity objects [7.0 $\lesssim$ 12+log(O/H) $\lesssim$
8.3] in comparison with the solar metallicity [12+log(O/H)$_{\odot}$=8.69; \cite{asplund09}], 
and actually include the most metal poor systems known in the 
local universe \citep[e.g.,][]{K06,W12}, for instance IZw18 with 12+log(O/H) $\approx$ 7.2  \citep[e.g.,][]{V98,P12}
and SBS~0335-052W for which 12+log(O/H) $\approx$ 7.0-7.2 \citep[e.g.,][]{I09}. Local metal-poor objects can be considered as
template systems that help us to understand distant star-forming
galaxies which cannot be studied to the same depth and angular resolution. 

Wolf-Rayet (WR) signatures (most commonly a broad feature centered at
$\sim$ 4680 \AA~ or blue bump), indicating the presence of WR stars
(the last observable stage during the evolution of massive stars
before the formation of neutron stars or black holes; e.g. \citealt{PC07}), have been found in the spectra of some HII galaxies
\citep[e.g.,][]{KS81,L97,G00,EPM10}. This is an important observational fact since single star, non-rotating stellar
evolution models fail in reproducing the WR content in low metallicity
environments \citep[see][and references therein]{BKD08}.  The
investigation of the WR content in galaxies is crucial to test stellar
evolutionary models, specially at low metallicities where more data
are needed to constrain these models.

This work is part of our program to investigate HII galaxies with WR
features using integral field spectroscopy (IFS; \citealt{K08,EPM11}). IFS has many benefits in a study of this
kind. Long-slit observations may fail in detecting WR features due to
their faintness with respect to the stellar continuum emission and
spatial distribution of WR stars across the galaxy. In particular in
low metallicity objects, like HII galaxies, the dilution of WR features and
the difficulty in spectroscopically identifying WR stars is even
stronger owing to the steeper metallicity dependence of WR star winds
which lowers the line luminosities of such stars \citep[e.g.,][]{CH06}. \cite{K08} demonstrated for the first
time the power of IFS in minimizing the WR bump dilution and finding
WR stars in extragalactic systems where they were not detected before
\citep[see also][]{C10,RGB10}.

This technique also lowers the difficulty when studying the spatial
correlation between WR stars and properties of the surrounding
interstellar medium (ISM; e.g., Legrand et al. 1997; de Mello et
al. 1998). Massive stars play a key role in the evolution of the
ISM. The mechanical energy injected by evolving stars through winds
and supernova (SNe) explosions can reach values $\sim$ 10$^{52}$ erg
\citep[e.g.,][]{M09}. The spatial distribution of gas near the
massive star clusters hosting WR stars may be affected by the
disruption of the ISM, which also determines how newly synthesized
metals are dispersed and mixed with the original gas from which the
stars formed.  Thus, a two-dimensional analysis of the ionized
material in HII galaxies helps us to better understand the interplay
between the massive stellar population and the ISM.  For instance
whether WR stars are a significant contributor to abundance
fluctuations on timescales of {\it t} $\sim$ 10$^{7}$ yr and to the
formation of high-ionization lines (e.g. He{\sc II}$\lambda$4686) are still
unsolved issues \citep[e.g.,][]{RK95,K11,SB12} that can be probed more precisely when applying IFS
to nearby galaxies like Mrk~178, as we will show in this
manuscript. The investigation on formation and thereabouts of
gamma-ray bursts and Type Ib/c supernova progenitors, believed to be
WR stars in metal-poor galaxies, may also benefit from the study
presented here \citep[e.g.,][]{WB06, modjaz08}.

The presence of a broadened 4686 \AA~feature, attributable to WR
stars, in the integrated spectrum of Mrk~178 had already been noticed
by  \cite{G88}. This system also appears in
the WR galaxy catalogue by \cite{S99} and is contained in
some studies of WR galaxies based on long-slit spectroscopy \citep[e.g.,][]
{G00,B05,BKD08}. In this work we present {\it the first two-dimensional
  spectroscopic study of the close-by HII galaxy Mrk~178}. At a distance
of 3.9 Mpc \citep{kar03} and with a metallicity $\sim 1/10
Z_\odot$, {\it Mrk 178 is one of the most metal-poor nearby WR galaxies}, which
makes it an ideal object for our project.

\begin{figure}
\center
\includegraphics[width=0.47\textwidth,clip]{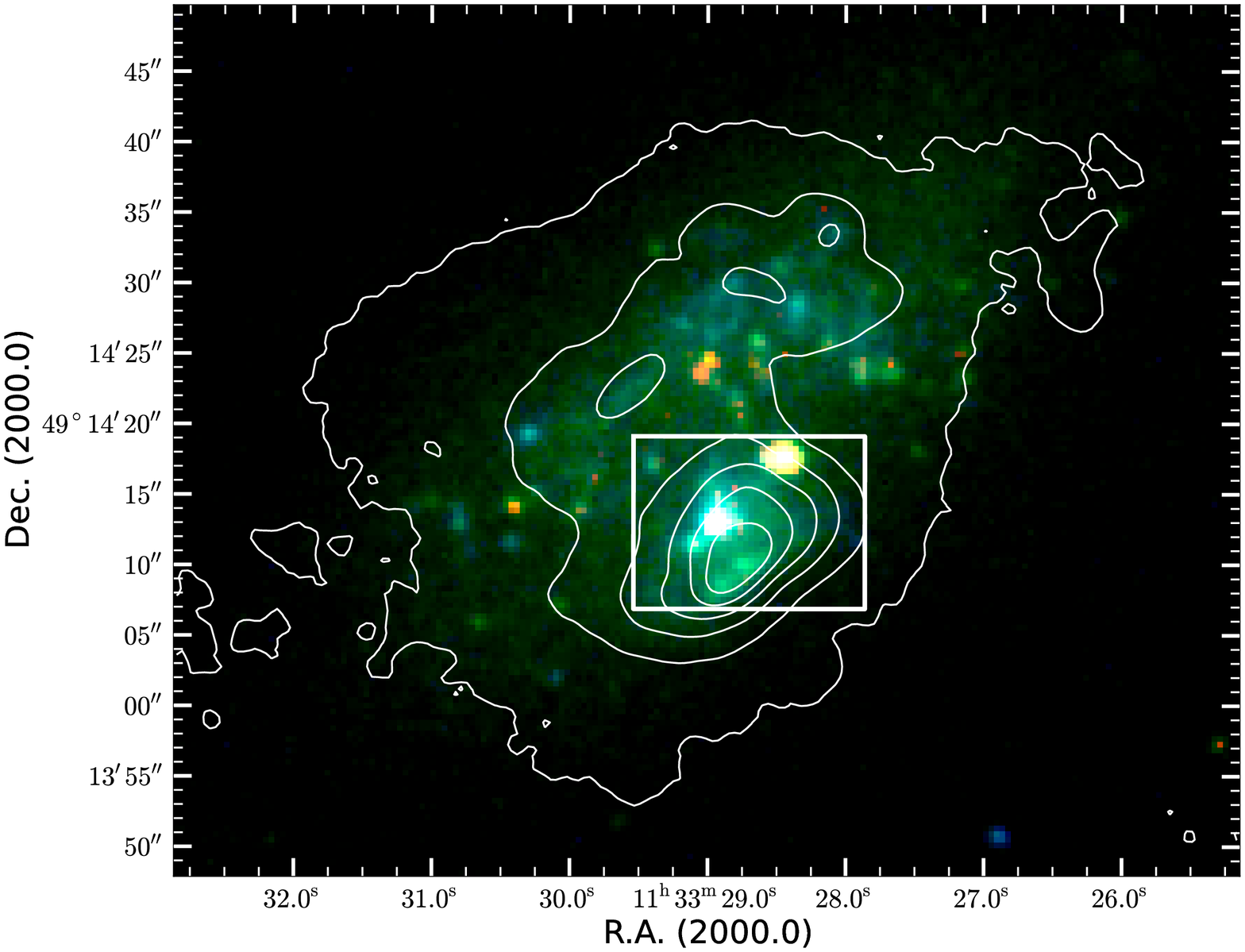}\\
\includegraphics[width=0.47\textwidth,clip]{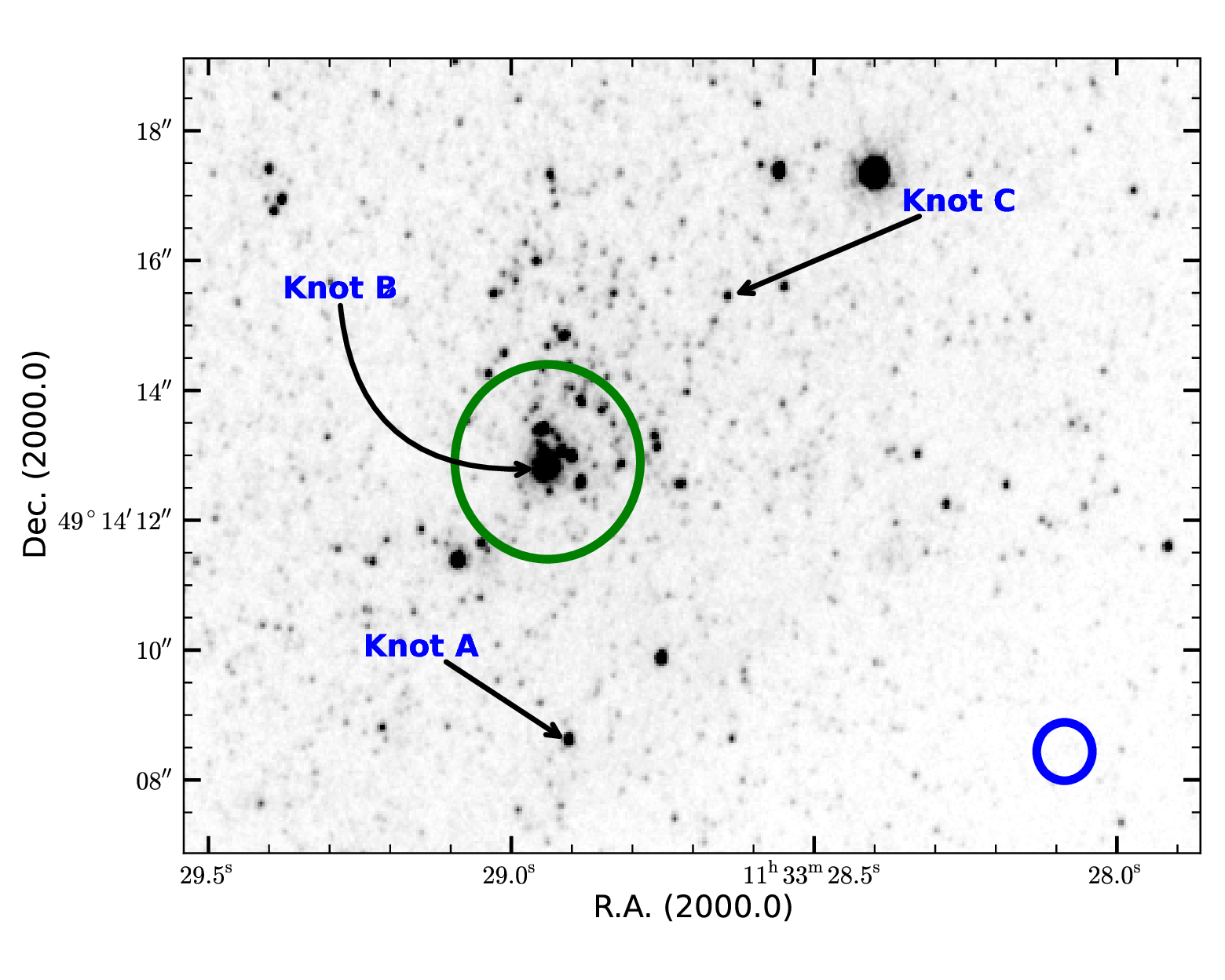}
\caption{{\it Top}: Three-band ({\it z,r,i}) SDSS color-composite image of
  Mrk~178 overlaid with the observed field of view (FOV) of
  WHT/INTEGRAL {\mbox  ($\sim $ 16~$\arcsec$ $\times$ 12$\arcsec$)} represented
  by the white box. The H$\alpha$ contours from Gil de Paz, Madore, \&
  Pevunova (2003) are also shown. {\it Bottom}: HST/WFC3 image of Mrk~178 in the continuum filter F547M,
  corresponding to WHT/INTEGRAL FOV. The big (green) circle 
  represents the approximate location of the SDSS fiber on
  Mrk~178. The small (blue) circle illustrates an individual spaxel (spatial element on the IFU) from the INTEGRAL
  fiber system. The three WR knots defined in
  section 3 are also marked. North is up and east is to the left.}
\label{mrk178} 
\end{figure}

We focus on investigating the spatial distribution of nebular
physical-chemical properties (e.g. electron temperature, gaseous metal
abundances) and excitation sources for the gas
in Mrk~178, and how these are related with the WR content derived
here. We discuss the significance of the observed spatial variations
of electron temperatures and chemical abundances found here, and the
origin of the nebular He{\sc ii}$\lambda$4686 emission.  Additionally, \cite{BKD08}
found Mrk~178 to be an intriguing object and significant outlier among
their sample of WR galaxies from  the Sloan Digital Sky Survey (SDSS;
\citealt{y00}). Based on the SDSS spectrum of Mrk~178, the authors
measured a remarkably high L(WR bump)/L(H$\beta$) $\approx$ 0.05. This
corresponds to a N(WR)/N(O) $\sim$ 0.5 which is much higher than
expected from any model at the low metallicity of Mrk 178. They
suggested that this could be caused by the limited aperture size
probed by the SDSS spectrum but were unable to test this
possibility. In this work we test this hypothesis using our integral
field unit (IFU) data.  Table~\ref{sample} compiles general properties of
Mrk~178.  In Fig.~\ref{mrk178} we show a three-color broad band image
from the SDSS of Mrk~178, and the HST/WFC3 image of Mrk~178 in the continuum filter F547M.

The paper is organized as follows. In Sect. 2, we report observations
and data reduction. The WR content and nebular properties across the sampled
regions of Mrk~178 are presented and discussed in Sect. 3.
Finally, Sect. 4 summarizes the main conclusions derived from this
work.

\begin{table}
\caption{General Properties of Mrk~178\label{sample}}
\centering
\begin{tabular}{lc} \hline
Parameter      &Mrk~178$^{a}$  \\ \hline
Other designation  & UGC~06541      \\
Morphological type$^{b}$ & iI,M dwarf   \\
R.A. (J2000.0)  & 11h 33m 28.9s         \\
DEC. (J2000.0)      & +49d 14' 14''        \\
redshift   &   0.0008            \\
D$^{c}$(Mpc) & 3.9   \\
Scale (pc/$\prime\prime$)     & 19 \\
(B-R)$^d$    &   0.89 $\pm$ 0.10         \\
{\it u}$^e$ (mag) & 15.133 $\pm$ 0.040   \\
{\it g}$^e$ (mag) & 14.228 $\pm$ 0.024 \\
{\it r}$^e$ (mag) & 14.139 $\pm$ 0.023  \\
{\it i}$^e$ (mag) & 14.131 $\pm$ 0.023  \\
{\it z}$^e$ (mag) & 14.153 $\pm$ 0.021  \\
A$_{V}$$^f$(mag) & 0.049   \\
R$_{25}$$^g$(arcmin) & 0.62 \\
SFR(H$\alpha$)$^h$(M$_{\odot}$ yr$^{-1}$) & 0.0085 \\
$\Sigma_{SFR}$$^i$(M$_{\odot}$ yr$^{-1}$ Kpc$^{-2}$) & 0.006 \\
\hline 
\end{tabular}
\begin{flushleft}
$^{a}$ Mrk~178 belongs to the Markarian lists of galaxies with strong
ultraviolet continuum \citep{mrk69}; $^{b}$ From \cite{gildepaz03}; $^{c}$ Distance based on the
tip of the Red Giant Branch distance from \cite{kar03};
$^{d}$ From \cite{gildepaz05}; $^{e}$ SDSS
photometry from \cite{SB12}; $^{f}$ Galactic extinction from
\cite{schlegel98}; $^{g}$ The optical radius R$_{25}$ (= D$_{25}$/2) in
arcminutes, from the RC3 catalog \citep{rc3}; $^{h}$ The
H$\alpha$ star formation rate from \cite{lee09}; $^{i}$ Star formation
rate per unit area [= SFR/($\pi$ $\times$ (R$_{25}$)$^2$)]. 
\end{flushleft}
\end{table}

\section{Observations and data reduction}\label{obs_datared} 

\subsection{Observations}
 
IFS data of Mrk~178 were obtained with the INTEGRAL fiber system
\citep{arribas98} in combination with the WYFFOS spectrograph
\citep{bingham94} at the 4.2 m William Herschel Telecope (WHT), Roque
de los Muchachos observatory. The data were taken with the 1024x1024
Tek6 CCD (Noise = 7 e; Gain = 2.1 e/ADU).  The observations were
performed on 2001 January 18 under photometric conditions and
atmospheric extinction in r' of $\sim$ 0.076 mag.

We used the R600B grating which covers from $\sim$ 3710 to 6750
\AA~and provides a full width at half-maximum (FWHM) spectral
resolution of $\sim$ 6\AA~and a linear dispersion $\sim$
3\AA/pixel. The data were taken with the standard bundle 2, for which the
field-of-view (FOV) is formed by 189 fibers (each $\sim$ 0.9$\arcsec$ in diameter) covering an area of
16$\arcsec$ $\times$ 12$\arcsec$ ($\sim$ 300 pc $\times$ 230 pc
at the distance of 3.9 Mpc) on the sky (see Fig.~\ref{mrk178}). One pointing of Mrk~178, encompassing its major star forming region
towards the south, was observed during 1.5 hours split into three
exposures of 1800 seconds each. The science frames were taken at
  airmasses $\lesssim$ 1.1 in order to avoid strong effects due to differential atmospheric refraction.

Additionally, calibration images (exposures of arc lamps and
of continuum lamps)  were obtained. Observations of the
spectrophotometric standard star Feige 34  were obtained during the
observing night to flux calibrate the data.

\subsection{Data Reduction}

The first steps of the data reduction were done through the P3d tool
which was developed to deal with integral-field spectroscopy data
\citep{sandin10}. After subtracting the bias level, the expected
locations of the spectra were traced on a continuum-lamp exposure.
The spectra were wavelength calibrated using the exposures of
CuAr+CuNe arc lamps obtained immediately after the science exposures.
We checked the accuracy of the wavelength calibration by measuring the
central wavelength of the [OI]$\lambda$5577 \AA~sky line in all fibers
and found a standard deviation of $\pm$ 0.20 \AA.

Fibers have different transmissions that may depend on the
wavelength. The continuum-lamp exposures were used to determine the
differences in the fiber-to-fiber transmission and to obtain a
normalized fiber-flat image, including the wavelength dependence.  In
order to homogenize the response of all the fibers, we divided our
wavelength-calibrated science images by the normalized fiber-flat
image. 

In the next step, the three exposures taken for the same pointing were
combined in order to remove cosmic rays, using the \textsc{imcombine} routine
in \textsc{IRAF}\footnote{IRAF is distributed by the National Optical
Astronomical Observatories, which are operated by the Association of
Universities for Research in Astronomy, Inc., under cooperative
agreement with the National Science Foundation}. Flux calibration was
performed using the \textsc{IRAF} tasks
\textsc{standard}, \textsc{sensfunc} and \textsc{calibrate}. We co-added the spectra of the
central fibers of the standard star exposure to create a 1D spectrum
that was used to obtain the sensitivity function. 
This sensitivity function was applied to the combined
  Mrk~178 science frame.

\begin{figure}
\center
\includegraphics[width=8cm,angle=0.0,clip]{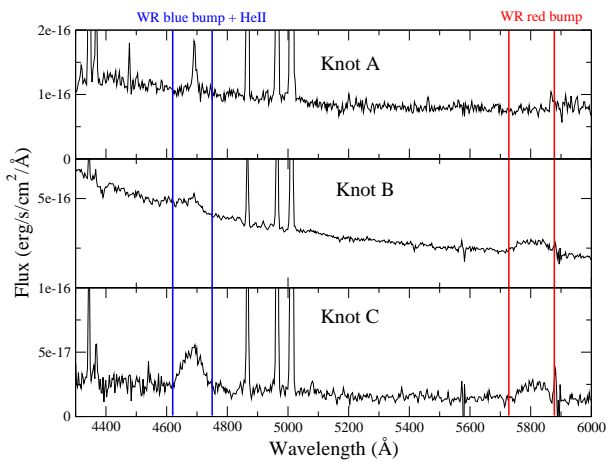}
\caption{Portion ($\sim$ 4300 - 6000 \AA) of the integrated spectrum, in units of erg s$^{-1}$ cm$^{-2}$ \AA$^{-1}$, for the three knots in which
the WR features are detected (see Figs.~\ref{maps},
\ref{oh} and \ref{no}). The spectral range for both blue and
red WR bumps are marked.}
\label{spectra_wr_knots} 
\end{figure}

\section{Results and Discussion}

\subsection{Number and location of WR stars}\label{wr_stars}

\begin{figure*}
\center
\includegraphics[width=10cm,angle=-90.0]{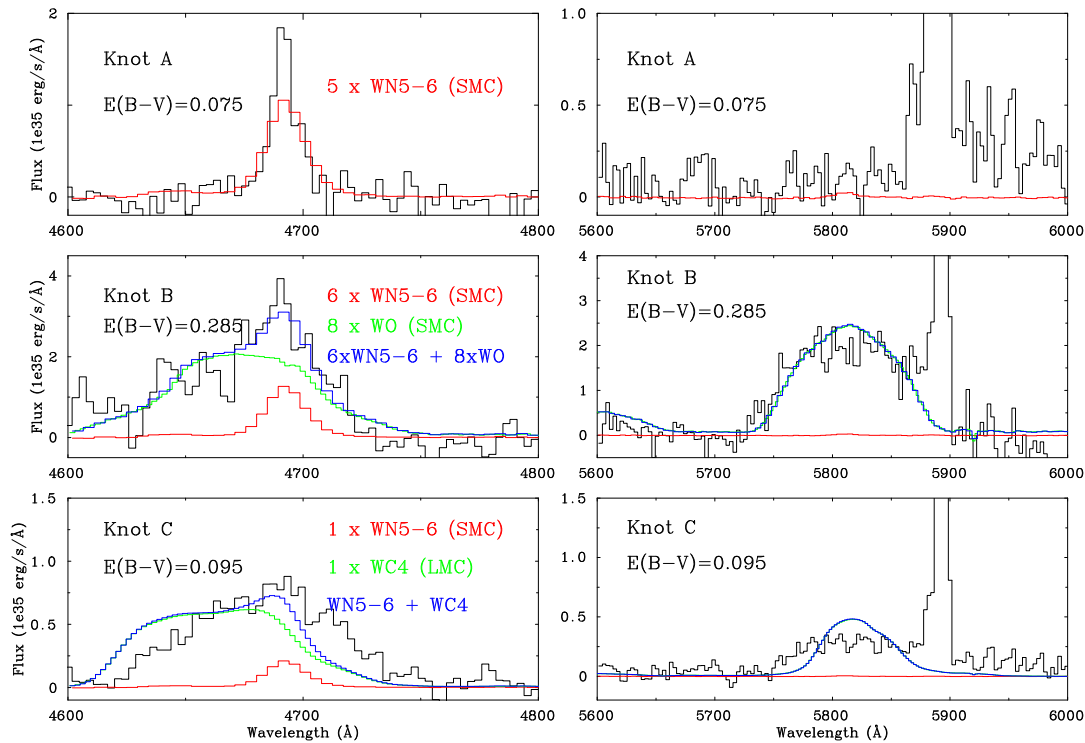}
\caption{From top to bottom: dereddened spectra of knots A, B, and C
  (black lines), corrected for a distance of 3.9 Mpc. Generic blue and
  red WR features [WN templates (red lines); WC/WO templates (green
  lines); composite WN+WO/WC templates (blue lines)] are shown for
  comparison. Templates of SMC WR stars are used where
  possible (WN,WO). We adopt WC4 LMC template since no SMC counterparts are available.
The spectra of the WR knots and WR star templates are continuum subtracted.}
\label{wr_templates} 
\end{figure*}

WR stars represent a late evolutionary phase of massive O stars whose
initial mass (M$_{ini}$) on the main sequence was, at solar
metallicity, $\gtrsim$ 25 M$_{\odot}$. At lower metallicities, this
M$_{ini}$ limit increases (e.g. Crowther 2007).  According to de Mello
et al. (1998), the M$_{ini}$ is $\gtrsim$ 90 M$_{\odot}$ for Z $\sim$
1/30 Z$_{\odot}$, based on non-rotating models for single
stars. Nevertheless, a significant number of WR stars detected in very
metal-poor objects appears contrary to the expectations of such models
(e.g. Izotov et al. 1997; Papaderos et al. 2006).  Including rotation
in the massive stellar evolution lowers the minimum mass required for
a star to enter the WR phase, increasing their population in
metal-poor environments \citep{mm05}.  WR stars are also extremely hot
and luminous with temperatures that sometimes exceed 50,000 K and some
with luminosities approaching $10^6 L_\odot$ \citep{PC07}. The spectra
of WR stars show strong, broad emission lines formed in their fast,
dense winds.  Two main classes of WR stars are defined based on
emission line ratios: nitrogen-type (WN) and carbon-type (WO and WC)
stars \citep[e.g.,][]{smith96,C98}. The so-called WN types are
nitrogen-rich helium stars that show the results of the CNO hydrogen
burning cycle \citep{smith73}. Besides the classic He-burning WN
stars, we also find in the literature examples of H-rich WR stars
usually very luminous\cite[WNh;][]{smith96,fmg,oli} which are believed
to be core-H burning \cite[e.g.][]{C95}. Neither H nor N are detected
in the spectra of carbon-type stars which show predominantly the
products of the triple-$\alpha$ and other helium burning reactions
\cite[see][]{PC07}.

Integrated spectra of star-forming galaxies sometimes reveal WR
spectral features, called WR bumps, which 
allow WR stars to be identified in both Local Group and more distant
systems. Exhaustive searches for extragalactic WR stars
have been the subject of several works \citep[e.g.,][]{B05,C07,NM11}. The blue
and red WR bumps observed in galaxy spectra are centered at $\sim$
4680 and 5808 \AA, respectively.  While WN stars are mainly
responsible for the formation of the blue bump, the carbon-type WR stars contribute
the most for the red bump emission.

WR features in the spectrum of Mrk~178 have been reported in previous
work \citep{G88,S99,G00}. From our IFU data we detect WR features in
three different regions in the galaxy which we defined as WR knots
(the location of the WR knots are marked on all maps of line
intensity, line ratio and ISM properties; see Figs.~\ref{maps},
\ref{oh} and \ref{no}). Knots A and B are composed by three
spaxels\footnote{ Individual elements of IFUs are often called 'spatial pixels' (commonly shortened to 'spaxel'); the term is used to differentiate between a spatial element on the IFU and a pixel on the detector.} while
knot C is represented by a single spaxel. The detection
of WR signatures at the locations of knots A and C is reported for the
first time in this work. We created one-dimensional spectra for knots
A and B by adding the emission from the corresponding spaxels.
Portions of the resulting spectra are shown in
Fig.~\ref{spectra_wr_knots} from which we can see the WR
signatures. The spectrum of knot A only displays the blue WR bump
while we see both blue and red WR bumps in knots B and C. 
No WR features were detected in any parts other than Knots A, B and C
across the sampled regions. We shall now
discuss the WR content in Mrk~178 within the INTEGRAL FOV.

According to Starburst 99 synthesis evolutionary models (Leitherer et
al. 1999) the expected number of WR stars for a star cluster at
  an age of $\sim$ 4 Myr (the age at
which models predict the WR content to be maximum), and
with the H$\alpha$ luminosity ($\sim$ 10$^{37}$ erg/s) and metallicity
similar to those measured for the three WR knots in Mrk178 is
$\lesssim$ 1.  State of the art models of stellar evolution that
  include the effects of rotation or binary evolution on the main
  sequence and beyond might be able to reproduce the observed WR
  features \citep[e.g.,][and references therein]{eldridge09}.
However, since full sets of tracks are not yet available and star
  forming regions do not fully sample the upper IMF, a more empirical
approach has been taken. In this regard, we will use template fitting
of individual WR stars to estimate the WR star content in this work.
This methodology has been successfully applied in synthesizing the WR
bumps in nearby spiral and irregular galaxies
\citep[e.g.,][]{sidoli96,HC06,moll07} including one example for
a metal-poor [12+log(O/H) $\lesssim$ 8.0] HII galaxy (IZw18;
\citealt{CH06}).

Here, to assess the number of WR stars in each knot, based upon currently
available templates, we use SMC
[12+log(O/H) $\approx$ 8.0; \cite{RD90}] template WR stars [early-type
WN (WN2-4), mid-type WN (WN5-6) from \cite{CH06}; WO from
\cite{kingsburgh95}] because of the low metallicity of Mrk~178
[12+log(O/H) $\approx$ 7.72; see section~\ref{variations}]. LMC
templates are used solely when SMC templates are unavailable (e.g. WC
subtypes). These template WR stars were adjusted for the redshift of
Mrk~178 (250 km\,s$^{-1}$). 

Figure~\ref{wr_templates} shows the comparison
between the observed WR bumps for our 3 knots and the WR template
spectra [predictions for WN stars (red), WC/WO stars (green) and
composite WN+WC/WO stars (blue)]. In view of the limited templates available
and variations in line strength and line width for individual WR stars we wish to
emphasise that the results presented here are not unique, but that we attempt
to match the morphology (line strength and width) using the fewest number of
subtypes. 

In Knot A, a broad blue ($\lambda_{c}$ $\sim$4686;
FWHM$\sim$15 \AA) emission feature is observed, together with possible
nebular He{\sc ii}$\lambda$4686 component. The FWHM ($\sim$5 \AA) of
this narrow He{\sc ii} component, comparable to that of the other
nebular emission lines (e.g., [O {\sc iii}]$\lambda$5007), and its
spatial extent are evidence of its nebular nature.  Applying the SMC
WN calibration from \cite{CH06} to the blue bump would require $\sim$
five equivalent WN5-6 stars.  

For knot B, we estimate that six WN5-6 and eight WO stars
[CIV$\lambda$5808 is very broad (FWHM$\sim$100 \AA) requiring a
WO instead of WC template] are needed to reproduce the WR features.
The blue WR spectral feature in knot C is fairly well reproduced by a
population of 1 SMC-like WN5-6 star and 1 LMC-like WC4 star (LMC
template WC4 stars were used since no WC4 have been observed in the
SMC). However the broad red feature requires a WO star to match the
observed morphology, while the blue comparison is much
worse. Alternatively, we checked whether a WN/C star\footnote{Our
  template, Brey 29 (BAT99-36), is the LMC composite WN4b/WCE \citep{foellmi03}} might enable a match to both blue and red
  features simultaneously, but a satisfactory solution could not be
  achieved since both features are much too broad with respect to a
  WN/C star.  While the small number of appropriate templates from
the SMC means that we cannot achieve a perfect fit for knot C, we find
that a population of $\sim 1$--$2$ WR stars is likely.


Of course, one needs to bear in mind that the numbers of WR stars implied from
these comparisons are only as good as the templates used. For
instance, there are only two carbon sequence WR stars at metallicities
below the LMC in the Local Group: the WO stars DR1 in IC~1613
\citep{kingsburghdr1} and Sand 1 in the SMC \citep{kingsburgh95}. In view of this, we have also examined the DR1 template,
but the width of the red bumps observed in knots B and C is in much
better agreement with the SMC template. In addition, the use of LMC templates
for WC stars (in the absence of SMC counterparts) is clearly not ideal. Still, the 
morphology of WC stars in the metal-poor dwarf IC~10 is reminiscent of
the LMC. Indeed, the average C\,{\sc iv} $\lambda\lambda$5801-12 line 
luminosity of IC~10 WC stars is consistent with that of the LMC WC population, in spite of
large uncertainties due to high foreground interstellar extinction \citep{Crowther03}.

\begin{figure}
\center
\includegraphics[bb=18 180 594 590,width=8cm,angle=0.0,clip]{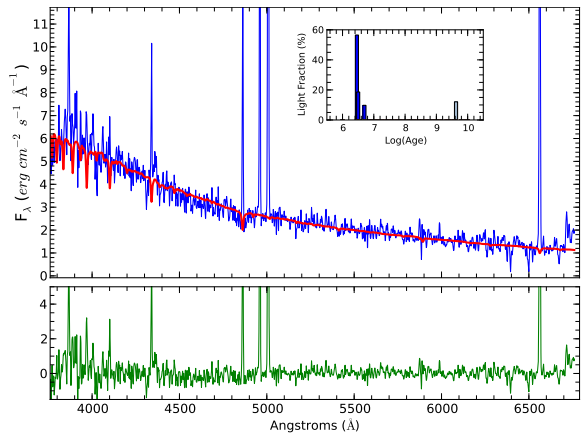}
\caption{This figure illustrates the main output from spectral synthesis for
  a single spaxel ($\sim$ 1$\arcsec$ from Knot B) in the region of
  Mrk~178 sampled in this work. The blue and red lines (top panel)
  correspond to the observed spectrum and the best-fitting synthetic
  SED for the stellar continuum, calculated with the population
  synthesis code STARLIGHT, respectively. By subtracting the latter
  from the former we obtain the pure nebular emission line spectrum
  (green in the bottom panel). The y-axis is in 10$^{-17}$ erg
  s$^{-1}$ cm$^{-2}$ \AA$^{-1}$. The smaller panel shows the age
  distribution of the stellar populations in the best-fitting synthetic SED  as a
  function of their luminosity (at 4200 \AA) in percent.}
\label{starlight} 
\end{figure}

Early-type WN templates would indicate a larger WR population (factor
of $\sim$ 10 times higher than for mid-type WN stars), although we
prefer mid-type WN stars throughout to early-WN stars because they
provide a better match to the He{\sc ii}$\lambda$4686 line width. Of
course, individual WN stars at a given subtype span a range of line
widths, but the range is modest for mid- and late-type stars
\citep[see Fig.2 from][]{CH06}.

In addition to the
observed line width constraint, mid-WN stars tend to be especially common
in giant HII regions across a range of metallicities, including our Milky Way
\citep[e.g., NGC~3603;][]{drissen95}, the LMC\footnote{Crowther \& Hadfield (2006) included 15 LMC
WN5-6 stars in their mid-WN template, of which 7 lie within 30 Doradus
(R144-147,R136a1,a2,a3).}\citep[e.g. 30 Doradus;][]{Moffat85}; and 
the SMC \citep[e.g. NGC 346;][]{Conti89}. Therefore, we might
expect these to dominate over early WN stars in knots A and B for
which the spatial extent is $\sim$ 4 arcsec ($\sim$ 80 pc at a
distance of $\sim$ 4 Mpc), the size of typical HII regions.  The size
of Knot C (radius $\sim$ 8 pc) is of the order of some ejecta nebulae
around isolated/binary WR stars \citep[e.g.,][]{EV92,stockbarlow10,alba12} which
supports our estimation of very few WR stars ($\sim$1-2) there.  It is
likely that the number of WR stars estimated here represents a lower limit
owing to the use of SMC templates and reduced line luminosities at
lower metallicities \citep{CH06}. Unfortunately, a more
quantitative assessment would require more individual metal-poor WR
stars to be included in low metallicity templates (IC10 is the only
suitable candidate at the present time).

\subsection{A 2D view of the warm ISM}\label{ISM} 

\subsubsection{Line fitting and map creation}\label{emlin}

To derive line ratios properly, we first subtracted from the observed
spectra the spectral energy distribution (SED) of the underlying
stellar population found by the spectral synthesis code
STARLIGHT\footnote{The STARLIGHT project is supported by the Brazilian
  agencies CNPq, CAPES and FAPESP and by the France–Brazil
  CAPES/COFECUB programme.} \citep{cid04,cid05,mateus06}. STARLIGHT
fits an observed continuum SED using a combination of the synthesis
spectra of different single stellar populations (SSPs) using a
$\chi^{2}$ minimization procedure.  We chose for our analysis the SSP
spectra from \cite{bruzual03}, based on the STELIB library of
\cite{leborgne03}, Padova 1994 evolutionary tracks and a
\cite{chabrier03} initial mass function between 0.1 and 100
M$_{\odot}$. We used the library metallicities (Z=0.001,0.004,0.008)
closest to the average oxygen abundance measured in the gas
[12+log(O/H) $\approx$ 7.72; see Section~\ref{variations}] and a set
of 38 ages from 1 Myr up to 10 Gyr. The reddening law from
\cite{cardelli89} with R$_V$ = 3.1 was applied. Bad pixels and
emission lines were excluded from the final fits. According to the
model SSPs, the correction factor for stellar absorption in the
EW(H$\beta$) is $\sim$ 10$\%$ on average. Our nebular lines have
EW(H$\beta$) $>$ 10 \AA~so the effect of the underlying stellar
population in Balmer lines involved in the reddening derivation is
negligible. The best spectral synthesis fits also show that young
stellar populations (age $<$ 10$^{7}$ yrs) dominate the light fraction
in most spaxels. An illustrative example of the observed,
modeled, and emission-line spectrum for a single spaxel in the FOV of
Mrk~178 is displayed in Fig.~\ref{starlight}.

\begin{figure*}
\center
\includegraphics[bb=1 1 505 364,width=0.45\textwidth,clip]{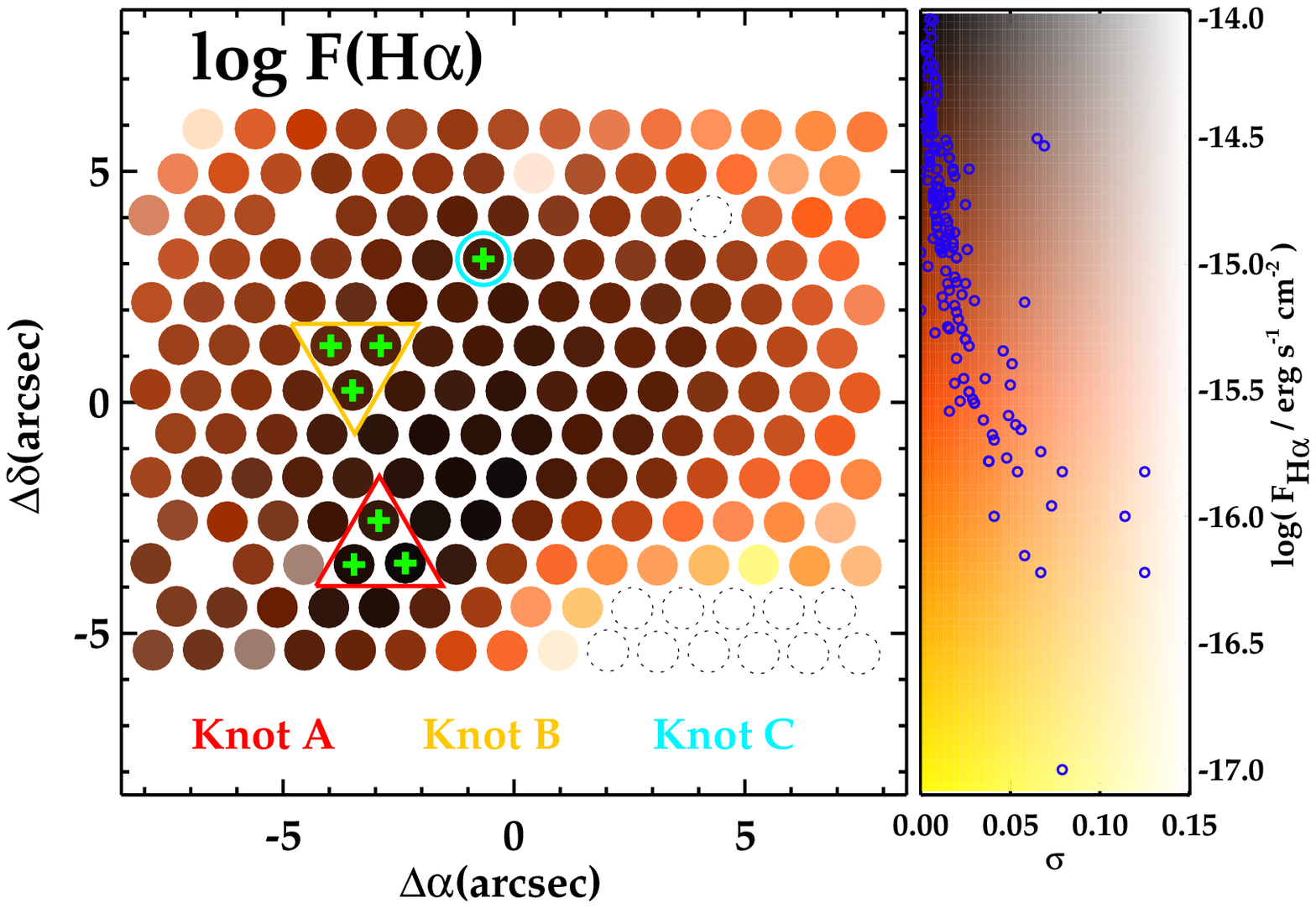}
\includegraphics[bb=1 1 505 364,width=0.45\textwidth,clip]{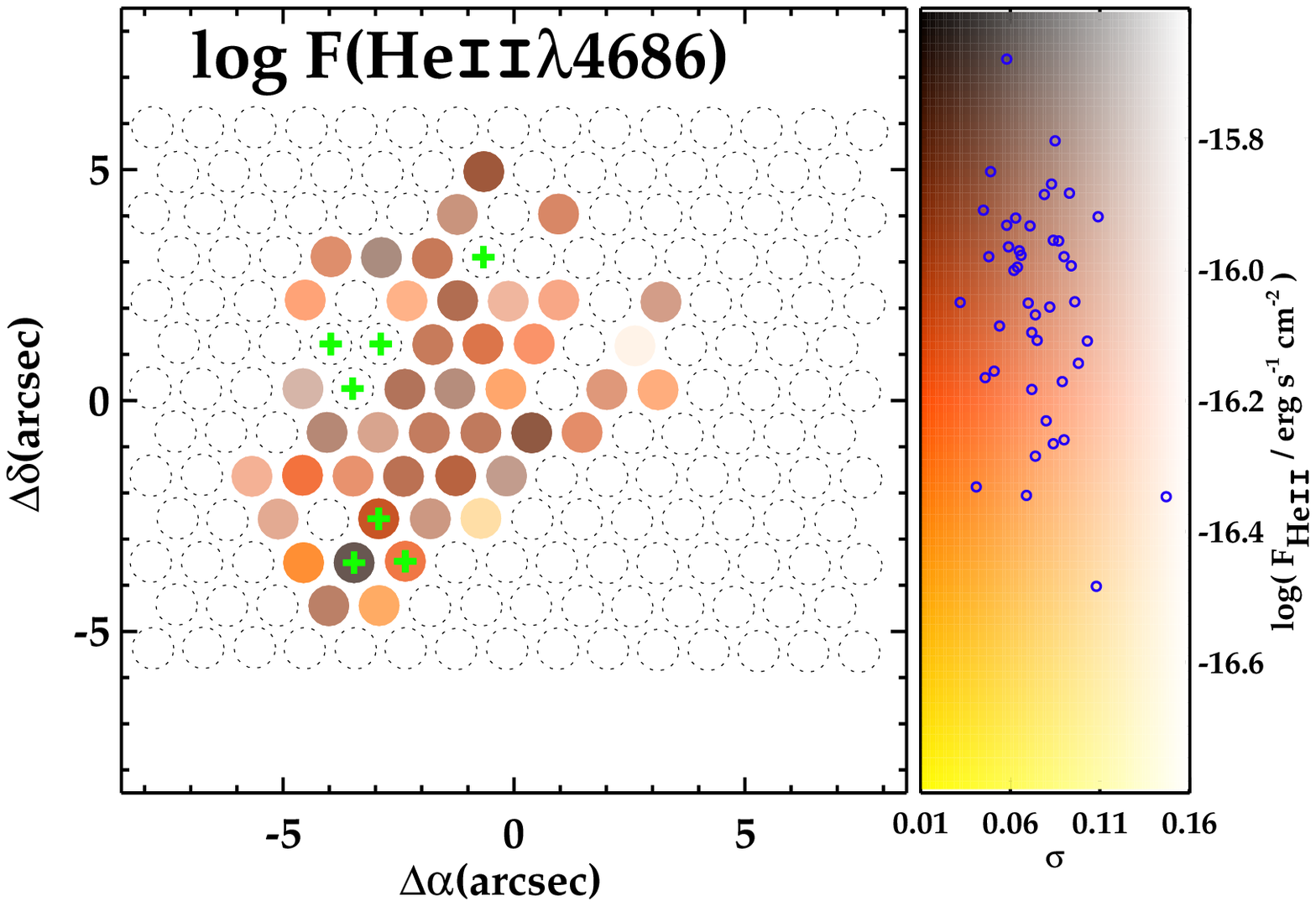}\\
\includegraphics[bb=1 1 505 364,width=0.45\textwidth,clip]{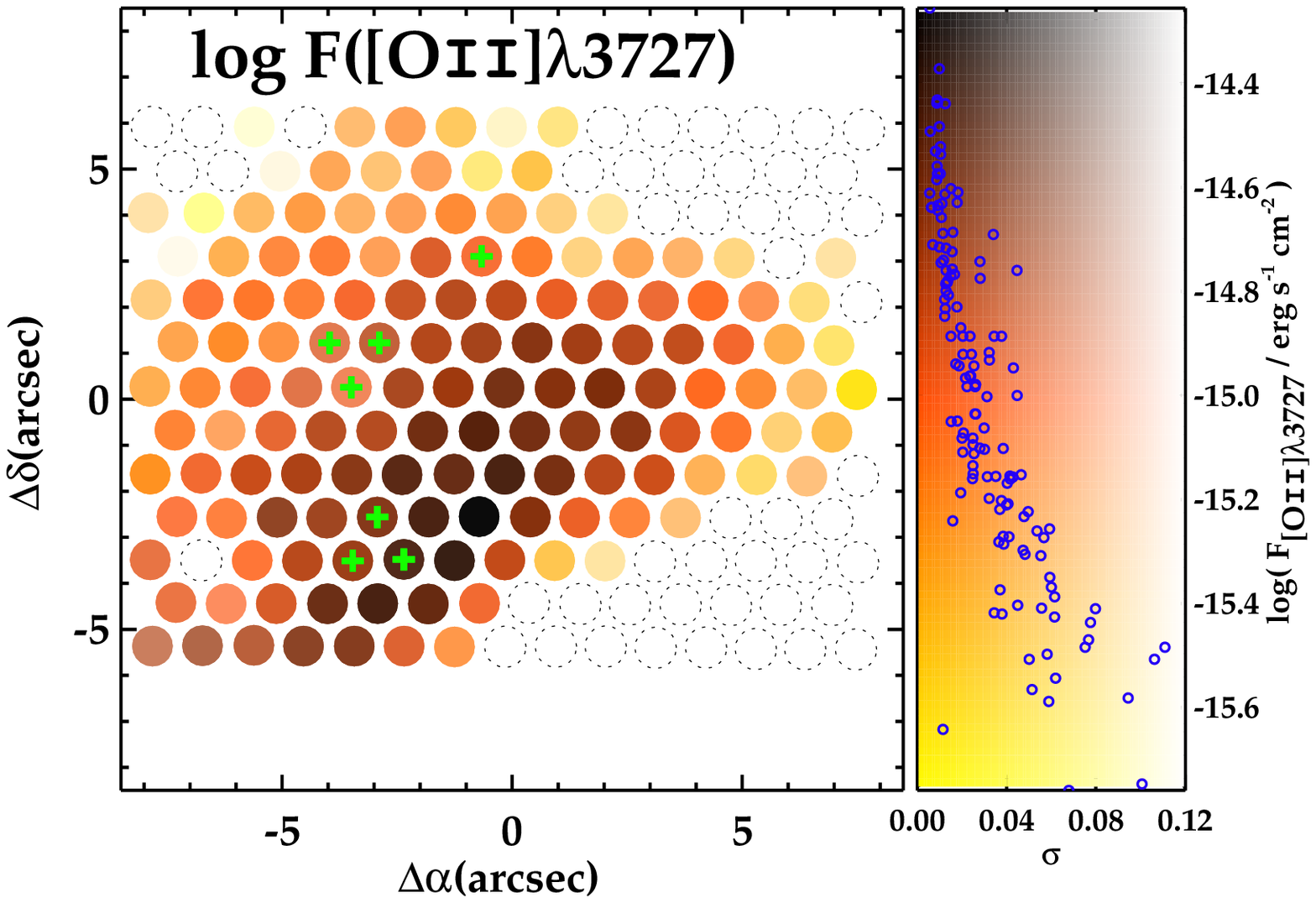}
\includegraphics[bb=1 1 505 364,width=0.45\textwidth,clip]{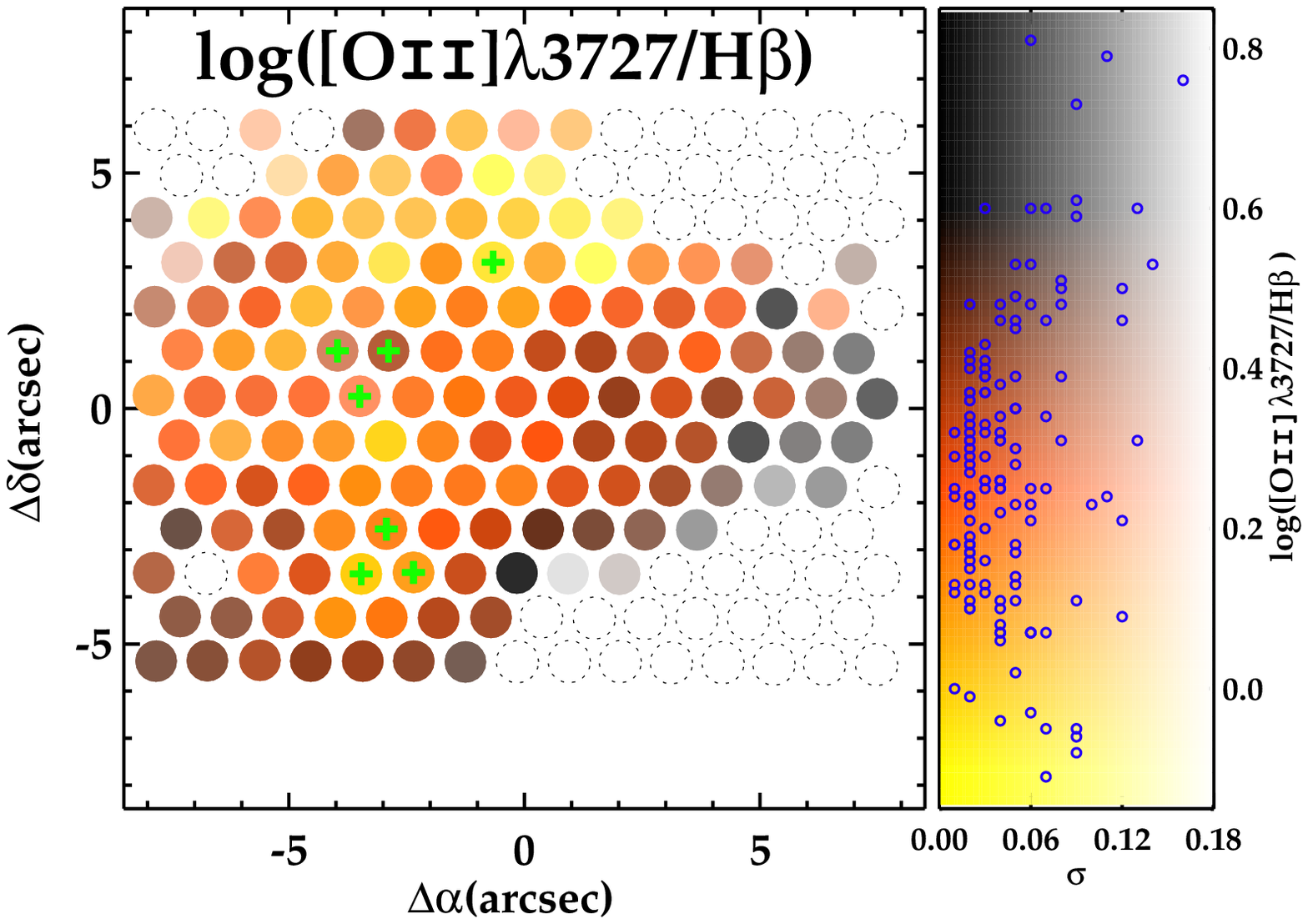} \\
\includegraphics[bb=1 1 505 364,width=0.45\textwidth,clip]{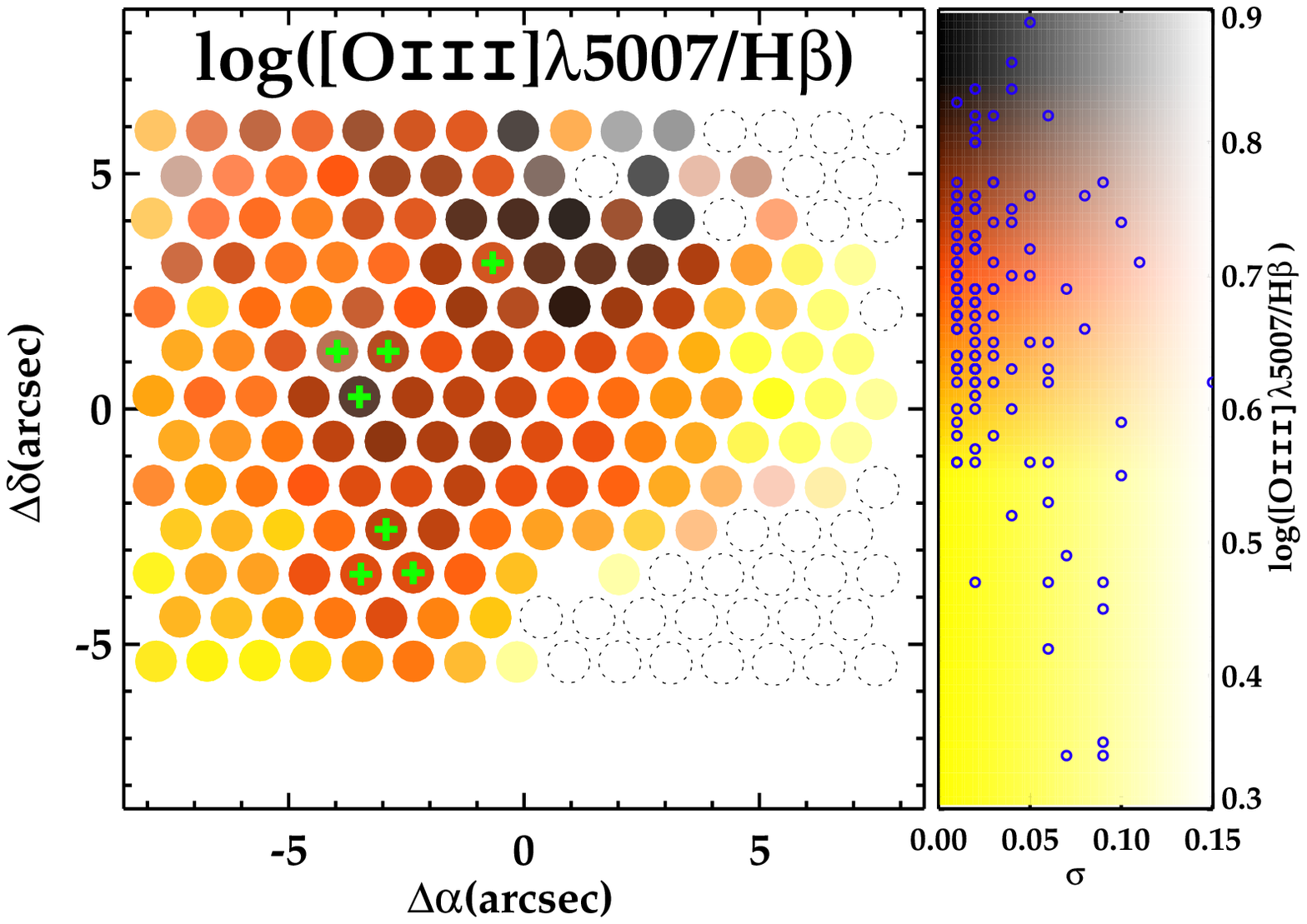}
\includegraphics[bb=1 1 505 364,width=0.45\textwidth,clip]{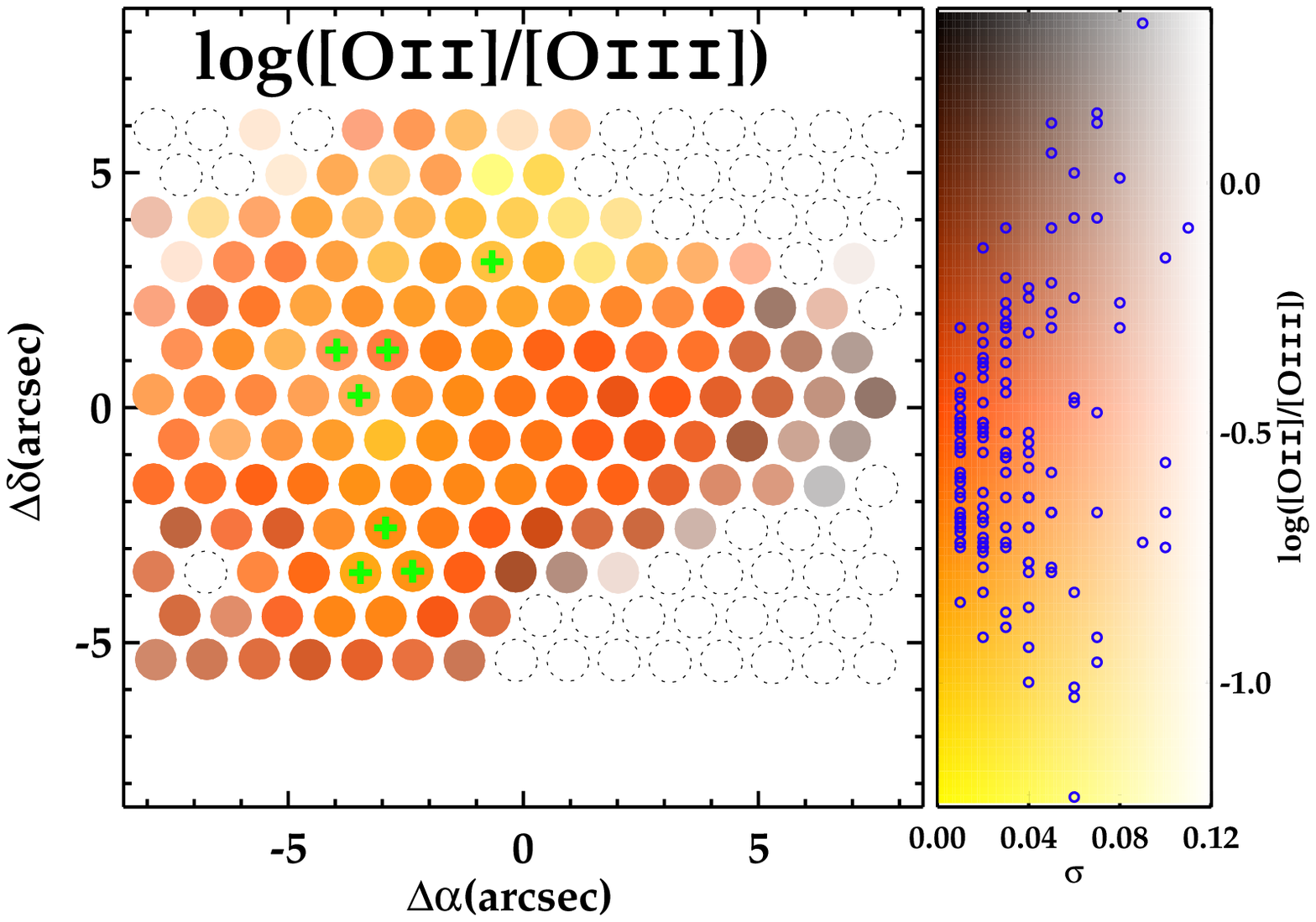}
\caption{Intensity and line ratio maps of Mrk~178: H$\alpha$, 
    narrow He{\sc ii}$\lambda$4686, [O{\sc ii}]$\lambda$3727, {\mbox [O{\sc ii}]$\lambda$3727/H$\beta$}, {\mbox [O{\sc iii}]$\lambda$5007/H$\beta$} and {\mbox [O{\sc
      ii}]$\lambda$3727/[O{\sc iii}]$\lambda$5007}. Dotted circles
  show the spaxels where no measurement is available. All maps are presented in logarithmic scale. The three WR knots
(A, B and C) are labeled on the H$\alpha$ flux map, and the spaxels where we detect WR features are marked with green crosses
on all maps. For each map the vertical axis of the right-hand side color bars
  represents the values of the corresponding line (-ratio) while the
  horizontal axis ($\sigma$) displays the uncertainties associated
  with each quantity, which are coded using an
  $\alpha$-channel (transparency of the spaxel color). According to the $\alpha$-channel coding, the more
  intense the color the lower the uncertainty. North is up and east to the left.}
\label{maps} 
\end{figure*}

After subtracting the underlying stellar population for each spaxel
spectrum, the emission-line fluxes are measured using
the \textsc{IRAF} task \textsc{splot}. The intensity of each emission line was
derived by integrating between two points given by the position of a
local continuum placed by eye. The line-flux errors were calculated using the
following expression: 

\begin{equation} 
\sigma_{line}=\sigma_{cont}N^{1/2}\left(1 + \frac{\rm EW}{N\Delta\lambda}\right)^{1/2}
\end{equation}
where $\sigma_{cont}$ is the standard deviation of the continuum near
the emission line, $N$ is the width of the region used to measure the
line in pixels, $\Delta\lambda$ is the spectral dispersion
in \AA/pixel, and EW represents the equivalent width of the line
\citep{rosa94}. This
expression takes into account the error in the continuum and the
photon count statistics of the emission line \footnote{We caution
    that for faint stars near the detection limit CCD
  readout noise can also be important.}. The relative
errors in the line intensities are $\sim$ 5-8$\%$ for the bright lines
(e.g. [O{\sc ii}]$\lambda$3727; [O{\sc iii}]$\lambda$5007;
H$\alpha$). Typical uncertainties for the weakest lines ([O{\sc
iii}]$\lambda$4363; He{\sc ii}$\lambda$4686; [N{\sc ii}]$\lambda$6584)
are about $\sim$ 15 $\%$ and may reach up to $\sim$ 20-30$\%$ in some
fibers. The final errors do not account for flat-field and
  instrumental response uncertainties which are not significant ($<$ 5$\%$).

\begin{figure*}
\center
\includegraphics[width=15cm,angle=0.0,clip]{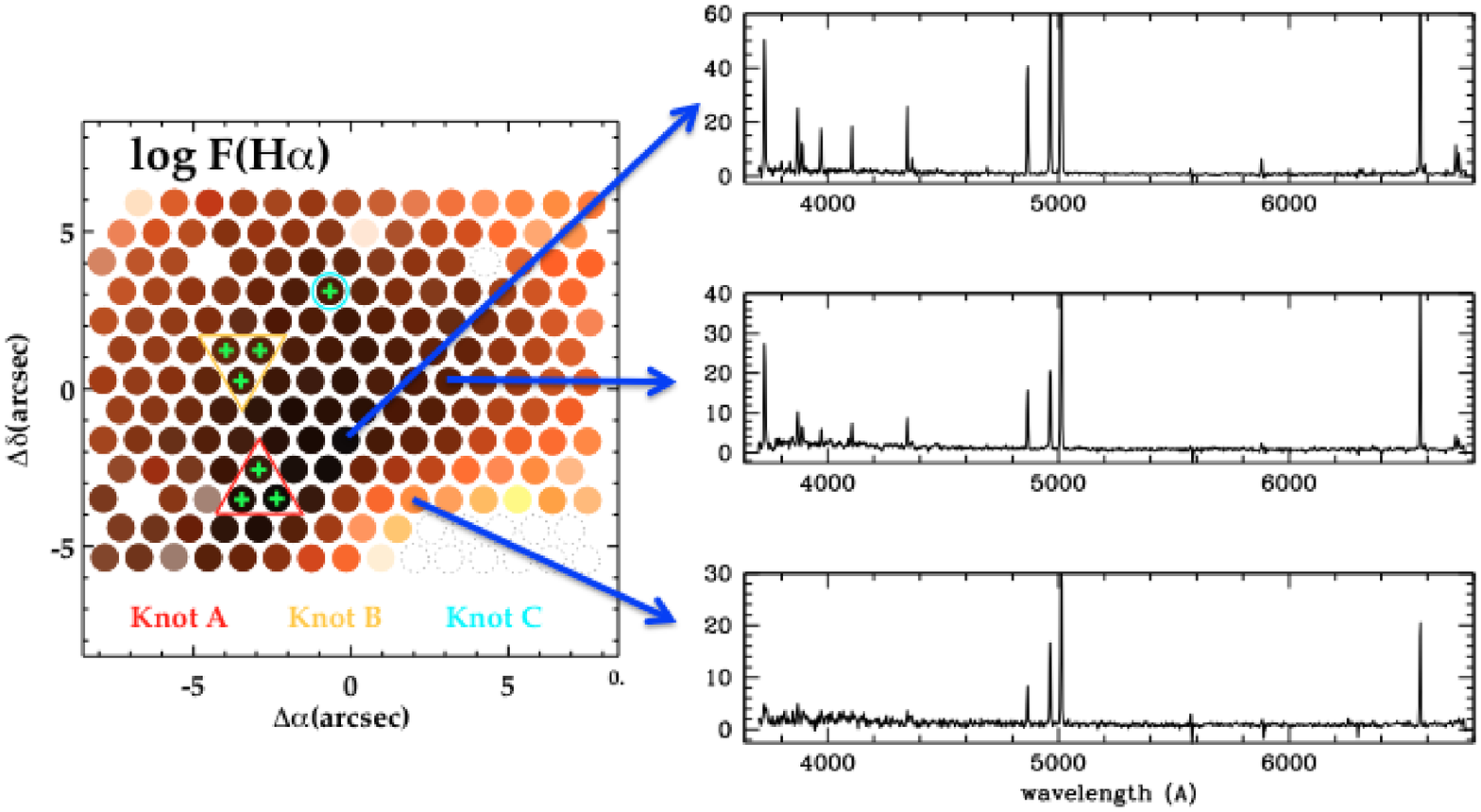}
\caption{Representative spectra in units of 10$^{-17}$ erg s$^{-1}$ cm$^{-2}$
  \AA$^{-1}$, corresponding to the nebular emission of three single
  spaxels across the FOV.}
\label{elspectra} 
\end{figure*}

Using our own IDL scripts we combine the flux intensities with the sky
position of the fibers to create the maps of emission lines and
line-ratios presented in this paper. Figure~\ref{maps} displays flux
maps of some of the relevant emission lines and line ratios. 
As a guide to the reader, the
spaxels where we detect WR signatures are indicated in all maps. The
H$\alpha$ and [O{\sc ii}]$\lambda$3727 maps show a larger area than
the He{\sc ii}$\lambda$4686 map because those lines are
among the brightest optical emission lines in our data sets. The maps 
of {\mbox [O{\sc ii}]$\lambda$3727/H$\beta$} and {\mbox [O{\sc iii}]$\lambda$5007/H$\beta$}
present opposite trends; {\mbox [O{\sc
ii}]$\lambda$3727/H$\beta$} peaks in the outer zones of the FOV while
the highest values of {\mbox [O{\sc iii}]$\lambda$5007/H$\beta$} are
observed close to the WR knot thereabouts where
the ionizing flux should be harder (see section~\ref{heii}  for
discusson on excitation sources). The map of the  {\mbox [O{\sc
    ii}]$\lambda$3727/[O{\sc iii}]$\lambda$5007} ratio which is an
indicator of the ionization parameter, is also shown. In
addition, Fig.\ref{elspectra} displays the H$\alpha$ map along with representative
emission-line spectra from spaxels in different positions across our FOV.

\subsubsection{Physical conditions and chemical abundances in the ISM}\label{ism_prop}

The reddening coefficient corresponding to each fiber spectrum,
c(H$\beta$), was computed from the ratio of the
measured-to-theoretical H$\alpha $/H$\beta$ assuming the
reddening law of \cite{cardelli89}, and standard HII region
characteristics ($T_{e}$ = 10$^{4}$ K; $n_{e}$ = 100 $cm^{−3}$) which
give  an intrisic value of H$\alpha $/H$\beta$ = 2.86 \citep{storey95}.

Using the de-reddened line-ratios, we derived the physical properties
and ionic abundances of the ionized gas for Mrk~178 following the
5-level atom FIVEL program \citep{shaw94} available in the
tasks \textsc{TEMDEN} and \textsc{IONIC} of the \textsc{STSDAS}
package.  We calculated the final errors in the derived
quantities by error propagation and taking into account errors in flux
measurements. Systematic errors are not included.

We obtained the electron densities, $n_{e}$, from the [S{\sc
ii}]$\lambda$6717/$\lambda$6731 line ratio. The derived estimates and
upper limits for $n_e$ place all of spaxel spectra in the low-density
regime ($n_{e}$ $\lesssim$300 $cm^{−3}$).

We derived the $T_{e}$ values of [O{\sc iii}] using the [O{\sc
iii}]$\lambda$4363/[O{\sc iii}]$\lambda$4959,5007 line ratio.  $T_{e}$[O{\sc ii}] was calculated
from the relation between [O{\sc ii}] and [O{\sc iii}] electron
temperatures given by \cite{EPM03} assuming an
electron density of 100 cm$^{-3}$.  We measured the faint auroral line
[O{\sc iii}]$\lambda$4363 for 60 spaxels with S/N $>$ 3. These spaxels cover a
projected area of nearly 42 arcsec$^{2}$ equivalent to $\sim$ 0.02
kpc$^{2}$, including the central starburst region. 

\begin{figure*}
\includegraphics[bb=1 1 505 364,width=0.44\textwidth,clip]{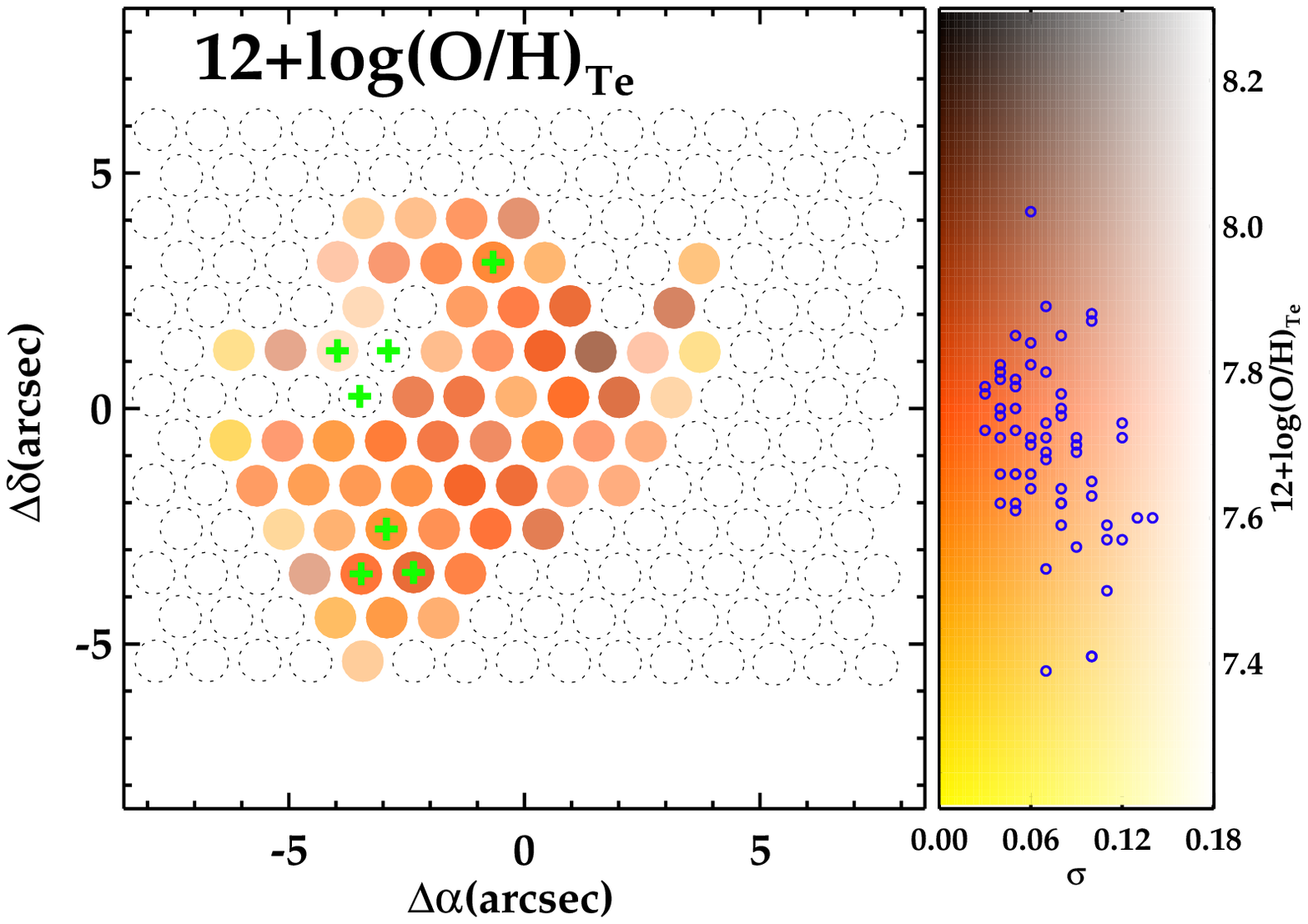}
\includegraphics[bb=1 1 505 364,width=0.44\textwidth,clip]{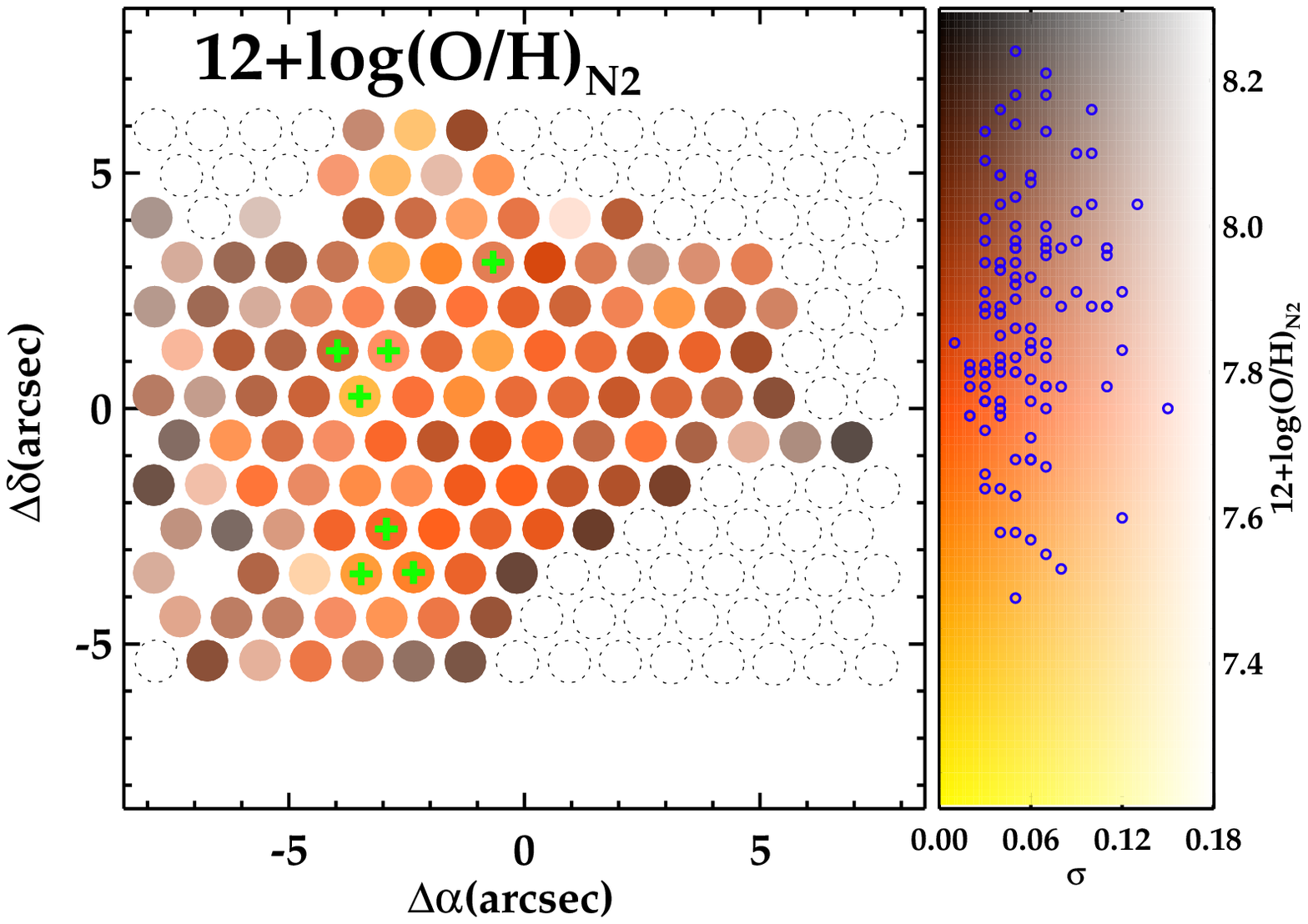}\\
\hspace{-0.50cm}\includegraphics[bb=24 59 544 560,width=0.41\textwidth,clip]{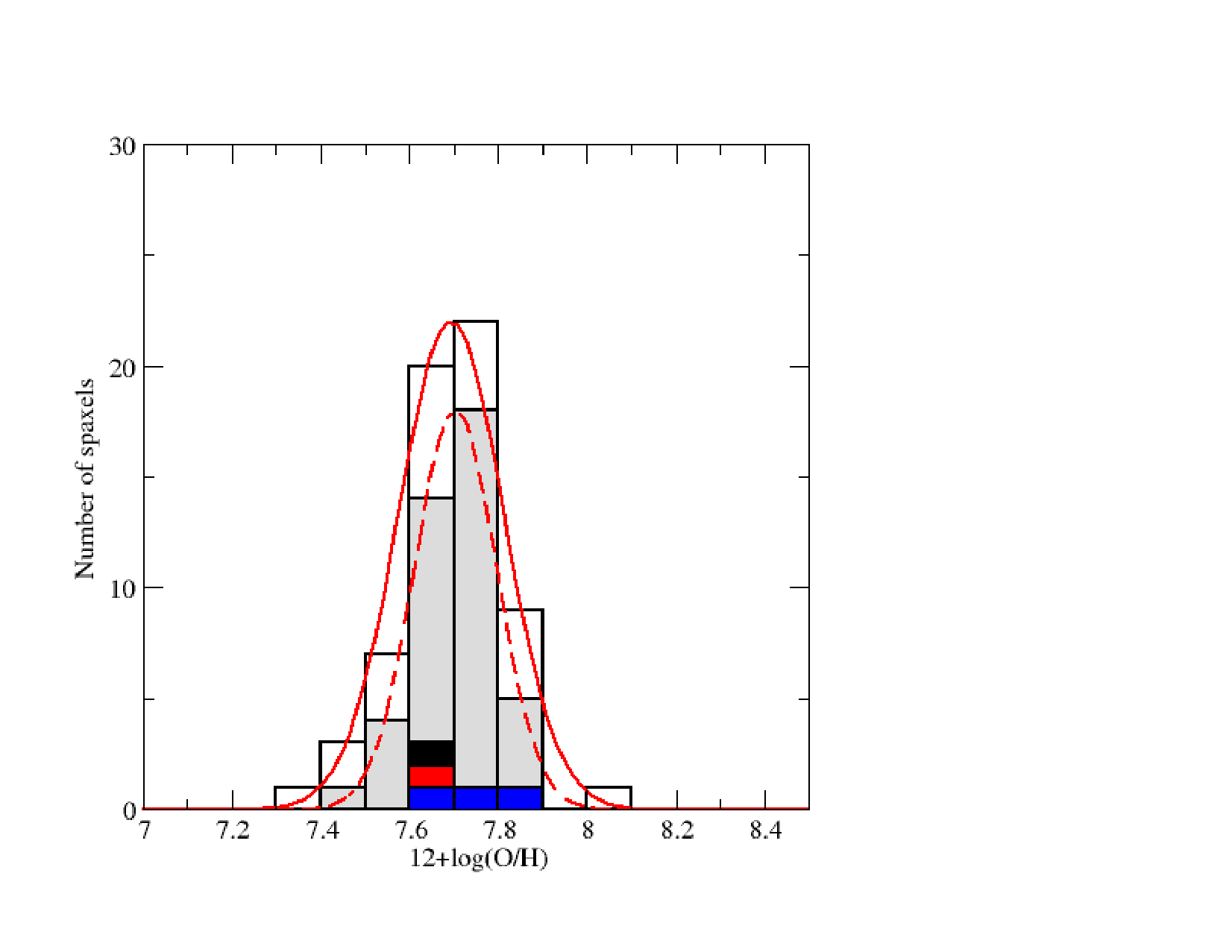}
\hspace{0.50cm}\includegraphics[bb=24 59 544 560,width=0.41\textwidth,clip]{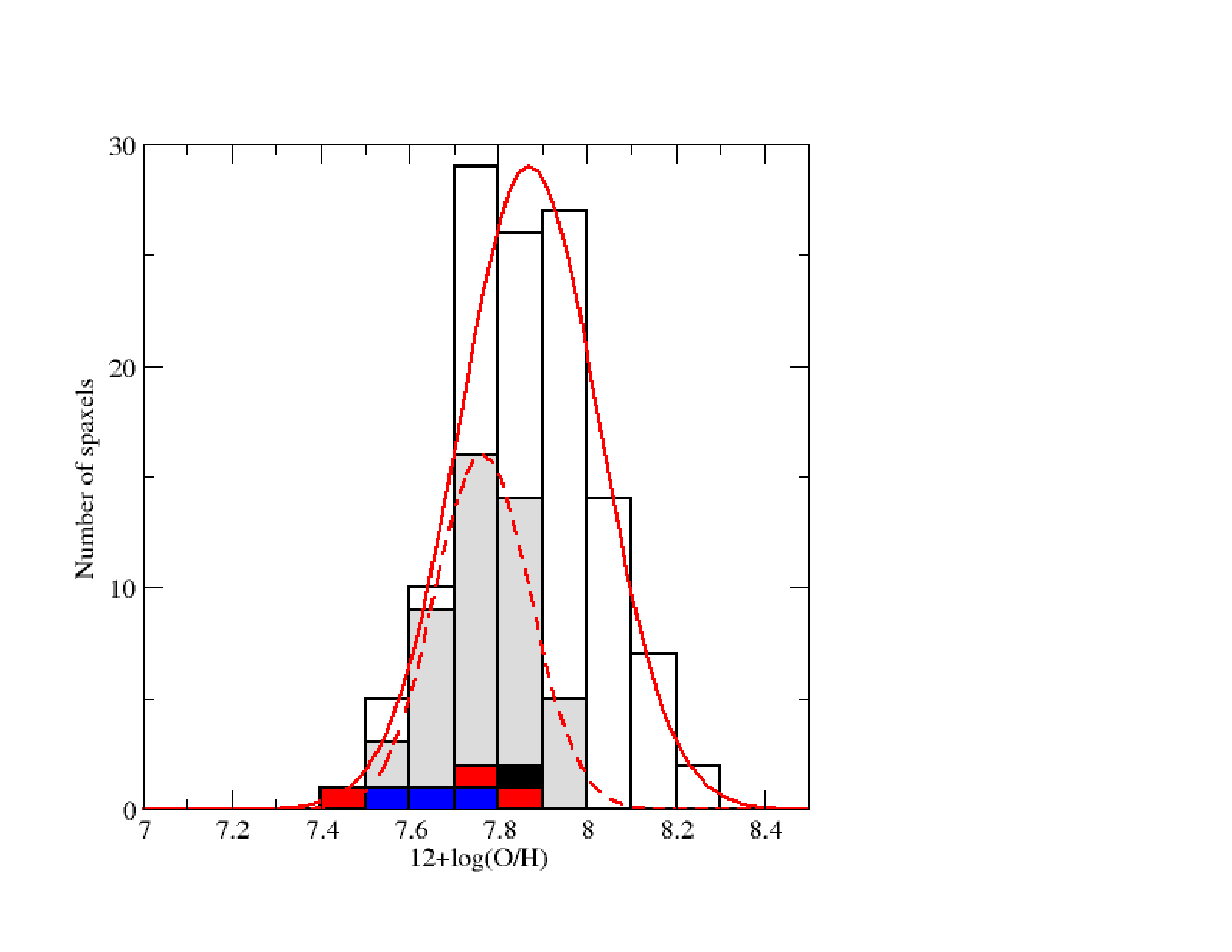}
\caption[caption]{{\it Top panels}: maps of oxygen abundance (12+log O/H) derived from $T_{e}$[O{\sc iii}] ({\it left}) and the
  N2 parameter ({\it right}); right-side color bars and green crosses are as indicated in Fig.~\ref{maps}. The corresponding histograms are in the {\it bottom row}. The gray bars indicate the spaxels where we
  measure the nebular He{\sc ii}$\lambda$4686, blue bars represent the spaxels
  showing only the blue WR bump (i.e. the spaxels within knot A; see Figs.~\ref{spectra_wr_knots} and~\ref{maps}), red bars
  indicate the spaxels within Knot B (see Figs.~\ref{spectra_wr_knots} and~\ref{maps}), all showing both the blue and red WR bumps, and the black bar represents the
  individual spaxel in the NW with both the blue and red WR bump
  detection (i.e. Knot C in Figs.~\ref{spectra_wr_knots}
  and~\ref{maps}). The solid and dashed curves are the
  Gaussian fits to all spaxels and only the spaxels where we detect
  nebular He{\sc ii}$\lambda$4686 line, respectively. The uncertainties of (O/H)$_{N2}$ are obtained by
    propagating the errors of [N{\sc ii}] and H$\alpha$ lines (involved in the N2
    parameter) in the derivation of O/H,
    and do not consider the typical  uncertainty associated with the N2
  parameter ($\sim$ 0.3 dex).}
\label{oh} 
\end{figure*}

The oxygen ionic abundance ratios, O$^{+}$/H$^{+}$ and
O$^{2+}$/H$^{+}$, were derived from the [O{\sc ii}]$\lambda$3727 
and \mbox{[O{\sc iii}]$\lambda\lambda$ 4959, 5007} lines,
respectively using the corresponding electron temperatures. The total
oxygen abundance is assumed to be: \mbox{O/H = O$^{+}$/H$^{+}$ +
O$^{2+}$/H$^{+}$}. With regards to the oxygen ionization correction
factor (ICF), a small fraction of O/H is expected to be in the form of
the O$^{3+}$ ion in high excitation HII regions when He{\sc ii}$\lambda$4686 emission
line is detected. In our case, we find the correction for the unseen
higher ionization stages of oxygen to be negligible; in the spaxels
where we measure He{\sc ii}$\lambda$4686, the ICF(O) correction would increase the
total O/H by less than $\sim$ 0.02 dex, which is considerably lower
than the typical error associated with our O/H measurements (see
top-left panel of Fig.~\ref{oh}). N$^+$ abundances were derived using the [N{\sc ii}]
emission line at 6584 \AA\ and assumming $T_{e}$[N{\sc ii}] $\sim$
$T_{e}$[O{\sc ii}]; the N/O abundance ratio was derived under the
assumption that \mbox{N/O = N$^{+}$/O$^{+}$}. Finally, based on the
He{\sc i}$\lambda$6678 and equations from \cite{olive04}, we
calculate He$^+$/H$^+$. These equations
were used for the appropriate values of $n_{\rm e}$, temperature, and
account for collisional deexcitation. We ignore
He{\sc i}$\lambda$5876 since it is affected by Galactic NaI absorption
and He{\sc i}$\lambda$4471 due to too low S/N.

We also derived O/H values based on the N2 parameter
\mbox{[=log([N{\sc ii}] $\lambda$6584/H$\alpha$)]} by \cite{EPMC09} to have an estimation of the metallicity over a
larger number of spaxels.  The N2 parameter is almost independent of reddening
effects, and the relationship between N2 and O/H is single-valued. 
Moreover, to derive N/O in a larger spatial
extent, we made use of the N2O2 parameter \mbox{[= log([N{\sc
ii}]$\lambda$6584/[O{\sc ii}]$\lambda$3727)]} which has a linear relation with
N/O at all ranges of metallicity
\citep{EPMC09}.

\begin{figure*}
\includegraphics[bb=1 1 505 364,width=0.44\textwidth,clip]{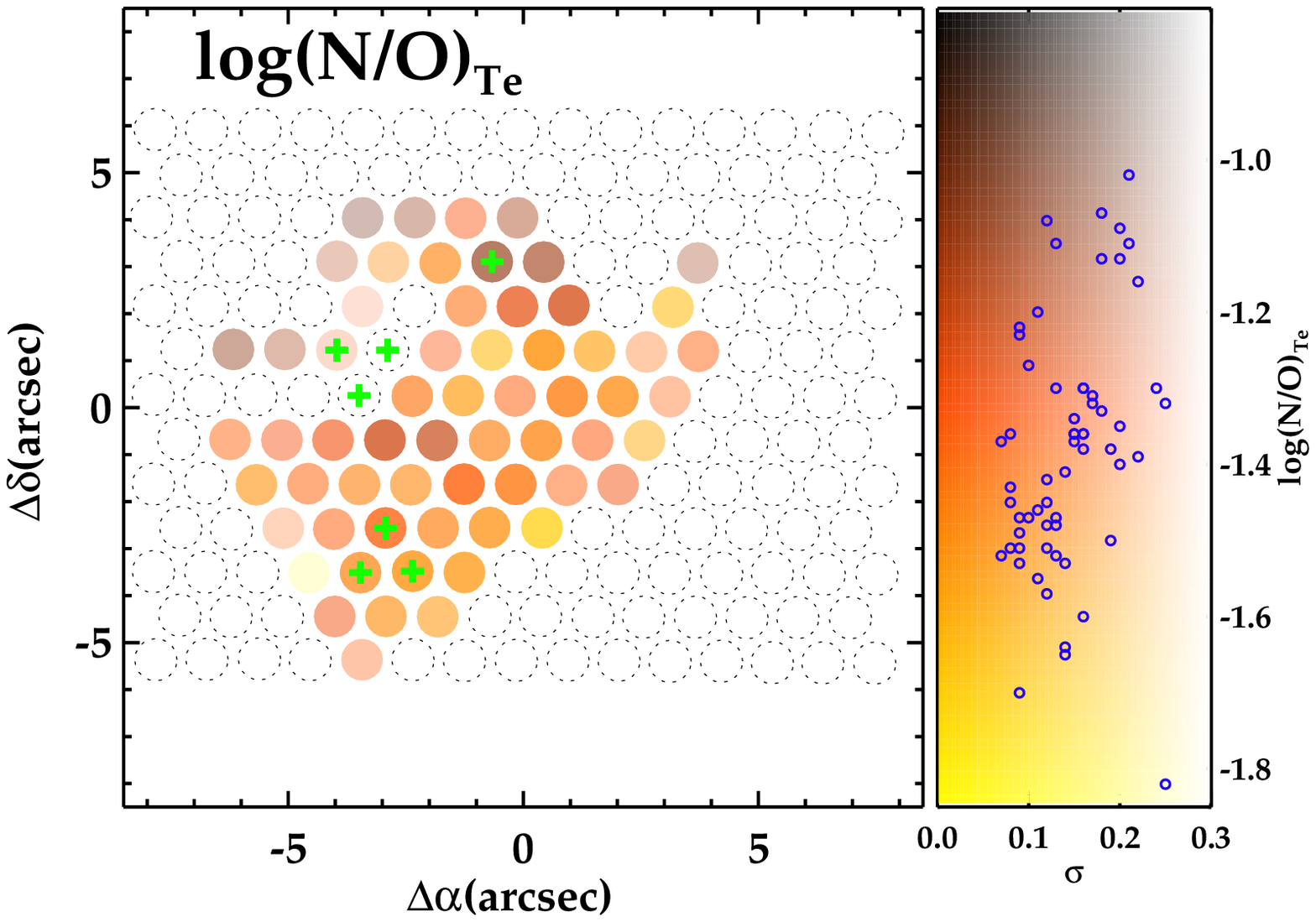}
\includegraphics[bb=1 1 505 364,width=0.44\textwidth,clip]{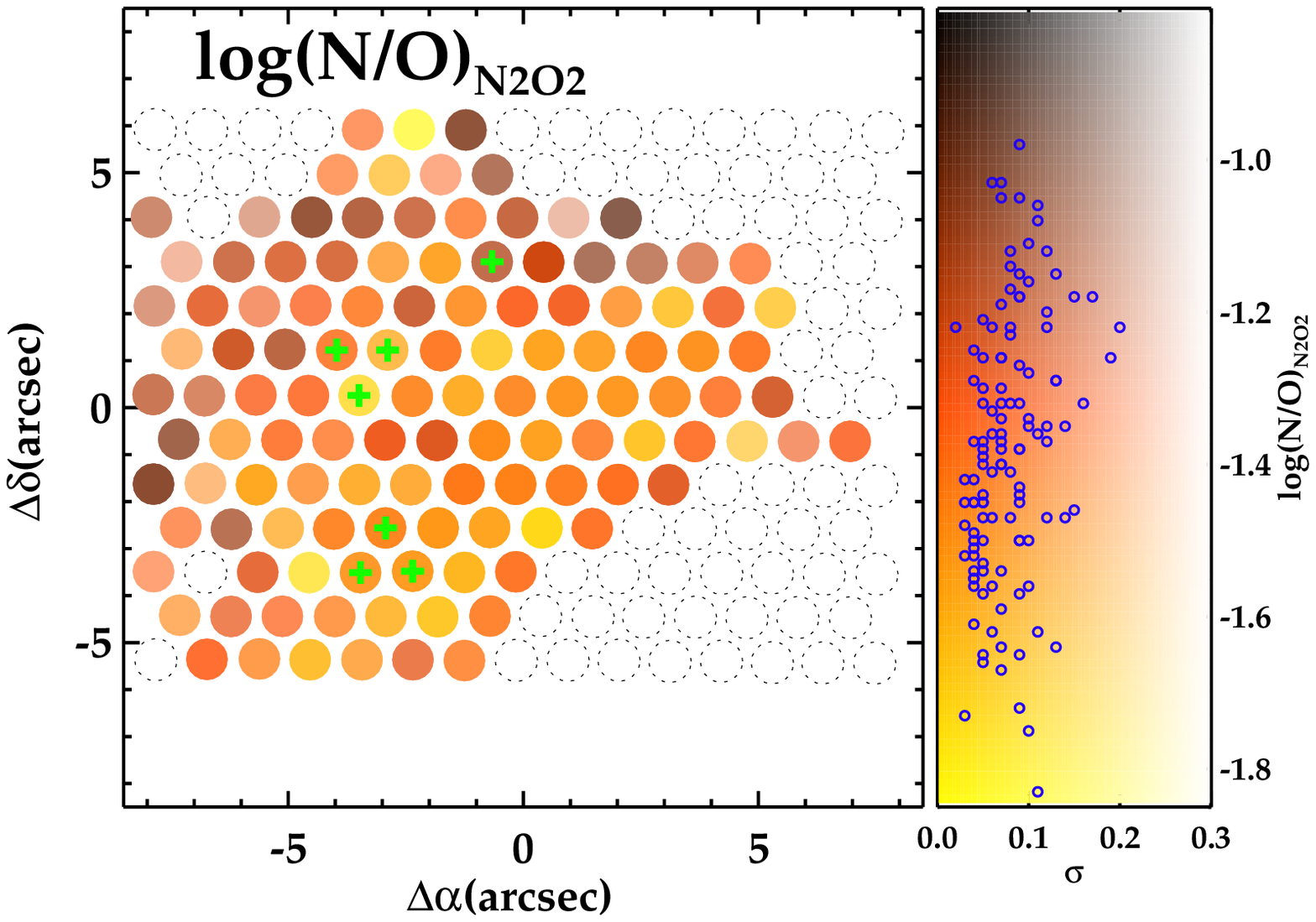}\\
\includegraphics[bb= 0 0 613 613,angle=0,width=0.42\textwidth,clip]{mrk178.fig8c.eps}
\includegraphics[bb=14 14 627 627,width=0.42\textwidth,clip]{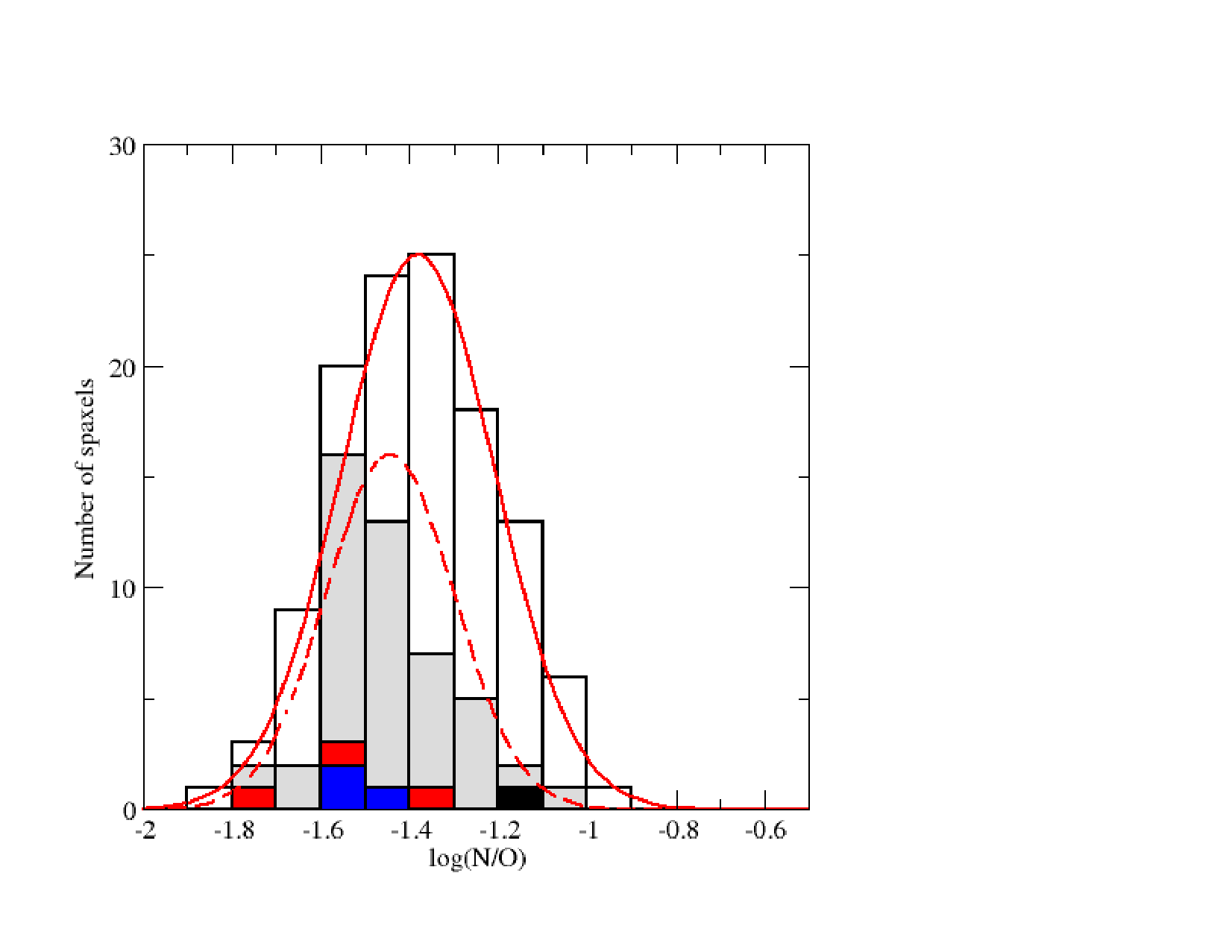}
\caption{{\it Top panels}: maps of log(N/O) derived from $T_{e}$[O{\sc iii}] ({\it left}) and the
  N2O2 parameter ({\it right}); right-side color bars and green
  crosses are as indicated in Fig.~\ref{maps}. The corresponding
  histograms are in the {\it bottom row}; the colouring of the bars
  and curves are the same as in Figure~\ref{oh}. The uncertainties of (N/O)$_{N2O2}$ are obtained by
    propagating the errors of [N{\sc ii}] and [O{\sc ii}] lines (involved in the N2O2
    parameter) in the derivation of N/O,
    and do not consider the typical  uncertainty associated with the N2O2
  parameter ($\sim$ 0.25 dex).}
\label{no} 
\end{figure*}

Maps and histograms of the derived O/H and N/O are displayed in
Figs.~\ref{oh} and \ref{no}, respectively.

\begin{table*}
\caption{De-reddened emission line-fluxes relative to 1000$\cdot$I(H$\beta$) 
and physical properties for the WR knots. The properties of the WR bumps are quoted.} 
\label{table_apertures}
 \centering 
\begin{minipage}{15.0cm}
\centering
\begin{tabular}{lccc}
\hline\hline 
Wavelength & Knot A & Knot B & Knot C    \\  \hline
3727 [O~II]   & 1320 $\pm$ 60  & 1670 $\pm$ 110 & 955 $\pm$ 60 \\ 
4363 [O~III]  & 135 $\pm$ 14    & 112 $\pm$ 17    & 147 $\pm$ 12 \\ 
4686 He~II    & 40 $\pm$ 4     & ---     & ---  \\
4861 H$\beta$ & 1000 $\pm$ 20  & 1000 $\pm$ 90  & 1000 $\pm$ 50 \\
5007 [O~III]  & 5290 $\pm$ 160 & 5213 $\pm$ 360 & 5330 $\pm$ 180 \\
6563 H$\alpha$ & 2810 $\pm$ 60 & 2810 $\pm$ 240 & 2770 $\pm$ 110 \\
6584 [N~II]    & 51 $\pm$ 14   & 93 $\pm$ 30   & 77 $\pm$ 16 \\
6678 HeI       & 37 $\pm$ 6    & ---    & 63 $\pm$ 16 \\
6717 [S~II]    & 112 $\pm$ 5   & 132 $\pm$ 11   & 129 $\pm$ 15 \\ 
6731 [S~II]    & 68 $\pm$ 5    & 89 $\pm$ 9     & 102 $\pm$ 16 \\  \hline
c(H$\beta$)    & 0.16 $\pm$ 0.03 & 0.26 $\pm$ 0.10  & 0.14 $\pm$ 0.05 \\
-EW(H$\beta$) (\AA)  & 80 $\pm$ 2 & 9 $\pm$ 1 & 33 $\pm$ 5 \\
log F(H$\beta$ (erg/s) & -14.00 & -14.25 & -14.75 \\  \hline
$n_{\rm e}$([S~II])(cm$^{-3}$)  & $<$10  & $<$10 & 210 \\
$T_{\rm e}$([O~III]) & 16186 $\pm$ 675 & 14600 $\pm$ 750 & 17079 $\pm$ 530 \\
12+log ($O^{+}/H^{+}$)  & 7.10 $\pm$ 0.08 & 7.29 $\pm$ 0.12 & 6.93 $\pm$ 0.08 \\ 
12+log ($O^{++}/H^{+}$) & 7.65 $\pm$ 0.05 & 7.76 $\pm$ 0.07 & 7.60 $\pm$ 0.03 \\
12+log ($N^{+}/H^{+}$)  & 5.65 $\pm$ 0.14 & 5.97 $\pm$ 0.17 & 5.81 $\pm$ 0.10 \\
12+log (O/H)$_{T_{\rm e}}$ & 7.76  $\pm$ 0.06 & 7.89 $\pm$ 0.08  & 7.69 $\pm$ 0.09 \\
log (N/O)$_{T_{\rm e}}$    & -1.45 $\pm$ 0.16 & -1.32 $\pm$ 0.20 & -1.12 $\pm$ 0.13 \\
He$^+$/H$^+$ ($\lambda$6678) & 0.11 $\pm$ 0.02  & --- & 0.18 $\pm$ 0.05  \\
He$^{2+}$/H$^+$ ($\lambda$4686)$^{a}$ & 0.004 $\pm$ 0.001 & --- & --- \\
\hline
log L(WR blue bump)$^{b}$  (erg/s) & 36.60 $\pm$ 0.07 & 37.52 $\pm$ 0.06 & 37.22 $\pm$ 0.07 \\ 
-EW (WR blue bump) (\AA)     & 7 $\pm$ 2        & 8 $\pm$ 2        & 50 $\pm$ 12 \\
log L(WR red bump)$^{b}$ (erg/s)   & --               & 37.88 $\pm$ 0.05 & 36.87 $\pm$ 0.06 \\
-EW (WR red bump) (\AA)      & --               & 20 $\pm$ 6       & 55 $\pm$ 15 \\  \hline 
\end{tabular} 
\end{minipage}
\begin{flushleft}
(a) estimated from
He{\sc ii}$\lambda$4686 and $T_{\rm e}$([O{\sc iii}]) \citep{I94}. 
(b) extinction-corrected luminosity of the WR bumps derived
at our assumed distance of 3.9 Mpc; to measure the WR features we
fitted the stellar continuum using a polynomical function and we
subtracted it from the WR emission. Nebular He{\sc ii}$\lambda$4686
line was removed before deriving the luminosity and EW of  WR
blue bump for knot A.
\end{flushleft}
\end{table*}

For the three WR knots A, B and C, the physical-chemical properties are
derived following the methodology described above and are presented in
Table~\ref{table_apertures}.

\subsubsection{Spatial variations of the ISM properties}\label{variations} 

Here, we refine the method described in \cite{EPM11} to study spatial variations of ISM
properties based on statistical criteria. We assume that a certain
physical-chemical property is homogeneous across our INTEGRAL FOV if
two conditions are satisfied: for the corresponding dataset (i) 
the null hypothesis
(i.e. the data come from a normally distributed population) of the
Lilliefors test \citep{li67} cannot be rejected at the 10$\%$
significance level, and (ii) the observed variations of the data distribution around the single
mean value can be explained by random errors;
i.e. the corresponding Gaussian sigma ($\sigma_{Gaussian}$) should be
of the order of the typical uncertainty of the considered
property; we take as typical uncertainty the square root of the weighted sample variance
($\sigma_{weighted}$). In Table~\ref{fits} we show the results from our statistical analysis
for relevant ISM properties.

For the electron temperature, despite the fact that the distribution
of its measured values can be represented by a Gaussian fit (based on the Lilliefors test), we find
that $\sigma_{Gaussian}$ $>$ $\sigma_{weighted}$ (see
Table~\ref{fits}). This suggests that as a first approximation 
random variables alone should not explain the T$_{e}$ distribution.
The source of these variations is hard to assess. A possible
explanation could be some error source that we do not take into
account (e.g. variable seeing). However if this was the explanation, this source of error
should affect all emission lines which does not seem to be the
case. Our statistical study indicates that the scatter in
T$_{e}$ can be larger than the scatter in O/H within
Mrk~178. \cite{V98} find similar results
for IZw18. The opposite is also found in the literature;
i.e. chemically homogeneous gas is found in star-forming regions with no
apparent T$_{e}$ variation \citep[e.g.,][]{rosa96}.

From Table~\ref{fits} we find that the measured values of O/H, both
from direct-method and N2 calibrator, are fitted by a normal
  distribution according to the Lilliefors test; additionally
$\sigma_{Gaussian}$ $\sim$ $\sigma_{weighted}$ in both methods. We
find the same results when considering only the O/H values from the spaxels
where we measure the nebular He{\sc ii}$\lambda$4686 (gray bars in the
histograms of Fig. ~\ref{oh}). Thus, according to our
statistical analysis, the oxygen abundance in the ISM of Mrk~178 is
homogeneous over spatial scales of hundreds of parsecs. However, we
must bear in mind that this statistical methodology ignores any
spatial information and cannot be used to discard the presence of
small-scale localized ($\sim$ 20 pc, our resolution element size)
chemical variations, as we will discuss further in this section.

Close inspection of the two histograms in
Fig.~\ref{oh} indicates that the Gaussian fit for the distribution of O/H$_{N2}$ is
slightly broader than that for the O/H$_{Te}$ histogram
($\sigma_{Gaussian,N2}$=0.16; $\sigma_{Gaussian,Te}$=0.12), and the
Gaussian peak is somewhat shifted towards
higher values. This probably reflects the dependence of N2 on the
ionization parameter.
Still, the mean values of the corresponding O/H$_{N2}$ and O/H$_{Te}$
distributions are consistent within the errors (see Table~\ref{fits}).

In the case of the N/O ratio, the width and peak of the histograms for
both (N/O)$_{Te}$ and (N/O)$_{N2O2}$ values are similar (see
Table~\ref{fits} and Fig.~\ref{no}). The distribution of N/O
  values can be represented by a Gaussian fit in agreement with the Lilliefors test, independently of the
method used to estimate the N/O ratio. Also, $\sigma_{Gaussian}$ is of
the order of $\sigma_{weighted}$ as is the case with O/H. Taking
  into account only the spaxels with nebular He{\sc ii}$\lambda$4686
  detection (gray bars in the histograms of Fig.~\ref{no}) yields
  the same results. So, overall, across the whole FOV ($\sim$ 300 pc x 230 pc) of
Mrk~178 we find no statistically significant variations in N/O. 

\begin{figure}
\center
\includegraphics[bb=1 1 560 544,angle=0,width=0.40\textwidth,clip]{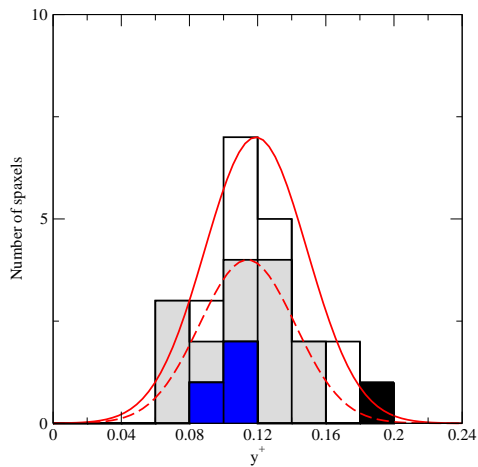}
\caption{Histogram of the He$^{+}$/H$^{+}$ ionic abundance ratio; the colouring of the bars and curves are the same as in Figure~\ref{oh}.}
\label{he_plus} 
\end{figure}

Since our statistical analysis does not consider any particular
spatial information, as mentioned above, to probe the existence of
small-scale ($\sim$ 20 pc) localized chemical pollution, we have
inspected the chemical abundances for the WR knots as derived in
section 3.2.2 (see Table 2).  From Table~\ref{table_apertures} and 
the N/O distributions in Fig.~\ref{no}, we note that WR knot C appears
to present a higher N/O.  Since N enrichment from WR star winds is
expected to come with He enhancement
\citep[e.g.,][]{pagel86,EV92,kob96}, we analyse the He$^{+}$/H$^{+}$
as well. From Fig.~\ref{he_plus}, the histogram of He$^{+}$/H$^{+}$,
and Table~\ref{table_apertures} one sees that WR knot C presents a
considerably high He$^{+}$/H$^{+}$ ($\sim$ 0.18) in comparison to the rest of the spaxels.  Thus, the
existence of a localized N and He\footnote{Considering the high ionization degree
 of the WR knots (O$^{++}$/O is
$>$ 0.70 for the three WR knots), their corresponding ICF(He) is expected to be $\sim$ 1 \citep[]{I07}} pollution by WR stars in knot C
cannot be ruled out. Since \cite{maeder92}, it is well known that WR
stars are significant contributors of He, N and C. At these
small-scales, chemical enrichment has been previously reported for
Galactic WR nebulae \citep[e.g.,][]{EV92,alba12} and for irregular
dwarf galaxies, for instance IC~10 \citep{lopez11} and NGC~5253
\citep[e.g.,][]{kob97,ana12}. A statistical detection of N
enhancement in galaxies showing WR features has been reported by
\cite{BKD08}. No evidence of chemical enrichment in the gas associated
with the WR knots A and B is found. Nevertheless, one should bear in
mind the detection of any chemical pollution on scales of $\sim$ few
parsecs will escape this work because of our resolution element size
($\sim 20$ pc at the distance of Mrk~178).

Summarizing, chemically homogeneous ionized gas on spatial scales of
$\gtrsim$ 100 pc appears to be a rule in HII galaxies
\citep[e.g.,][]{lee04,K08,EPM09,C09mrk409,C09mrk1418,EPM11,RGB12}.  One may speculate
that the ionized ISM in Mrk~178 and other HII galaxies appears
presently well homogenized because the freshly synthesized elements
were rapidly dispersed, on timescales $<$ 10$^{7}$ yr, and mixed very
quickly in the ISM. This hypothesis requires that ejected metals are
both homogeneously dispersed and mixed with the ambient ISM on time
scales comparable to that of the HII regions which seems to be
unlikely based on hydrodynamical and physical processes in the gas
\citep[see][]{RK95,T96,KS97}. Alternatively, the scenario proposed by
\cite{T96}, in which the ejecta from stellar winds and supernovae
could undergo a long ($\sim$ 100 Myr) cycle in a hot phase ($\sim$
10$^{6}$ K) before mixing with the surrounding ISM, has been discussed
in the literature \citep[e.g.][]{KS97,vanZ06} as a possible
explanation for the homogeneous chemical appearence of HII galaxies on
spatial scales of $\gtrsim$ 100 pc.  However the fate of the metals
released by massive stars in HII regions is still an open
question, and  the processes of metal dispersal and mixing are very
difficult to model \citep[see][]{recchi03,recchi13}. Thus, to better understand how the
ejected metals cool and mix with the ISM, further investigation on the
metal content of the different ISM phases is needed but is beyond the
scope of this work \citep[e.g.][]{lebouteiller09}.

In this work we find that the representative metallicity of Mrk~178 is
12+log(O/H) = 7.72 $\pm$ 0.01 ($\sim$ 1/10 of the solar metallicity)
which represents the derived error-weighted mean value of
O/H and its corresponding statistical error
from all the individual spaxel measurements obtained from the
direct-method (for this O/H distribution,  the square root of the weighted sample variance
associated is 0.13 dex; see Table~\ref{fits}).

\begin{table*}
\centering
\caption{Results from the statistical analysis for the distributions
  of  the physical conditions and chemical abundances in the ISM across the INTEGRAL FOV of Mrk~178.}
\label{fits}
\begin{minipage}{9.0cm}
\begin{tabular}{lccccc}
\hline\hline \\
& \multicolumn{5}{c}{Statistical properties} \\
ISM property & $\mu_{weighted}$$^{a}$ & $\sigma_{weighted}$$^{b}$ & $\mu_{Gaussian}$$^{c}$ & $\sigma_{Gaussian}$$^{d}$ & Sign.(\%)$^{e}$  \\
& & & & &\\ \hline
\multirow{2}{*}{log([O{\sc ii}]/[O{\sc iii}])} & & & & &  \\ 
 & -0.56  & 0.33  & --- & --- & $<$ 10\\ \hline
\multirow{2}{*}{n([S{\sc ii}]) (cm$^{-3}$)} & & & & &  \\ 
 & 110   & 95  & --- & --- & $<$ 10\\ \hline
\multirow{2}{*}{T([O{\sc iii}]) (K)} & & & & &  \\  
& 16770 & 760 & 17885 & 2200 & 26 \\ \hline
12+log(O/H) & & & & &      \\
(direct method)  & 7.72  & 0.13 & 7.69 & 0.12 & 80   \\ \hline 
 12+log(O/H)  & & &  &  &    \\
 (N2 method)  & 7.84 & 0.16 & 7.87 & 0.16  & 43  \\ \hline
log(N/O) &  & & & &   \\
(direct method)& -1.40  & 0.17 & -1.34 & 0.16 & 37    \\ \hline                        
log(N/O)      & & & & &  \\
(N2O2 method) & -1.42 & 0.19 & -1.38 & 0.17 & 75   \\ \hline 
\end{tabular}
\end{minipage}
 \begin{flushleft}
(a) error-weighted mean;(b) square root of the weighted sample variance associated with
the corresponding weighted mean; (c) mean of the Gaussian
distribution; (d) standard deviation of the Gaussian distribution; (e)
significance level of the null hypothesis in the Lilliefors test. We assume that the distribution of a certain property can be represented by a Gaussian fit if the null hypothesis is not rejected (i.e. Sig. $>$ 10\%)
\end{flushleft}
\end{table*}


\subsubsection{Excitation sources and nebular He{\sc ii}$\lambda$4686 emission}\label{heii}

Although photoionization by massive stars is the most likely main
excitation source in HII regions, some questions still remain
regarding the origin of high-ionization nebular lines as for instance
He{\sc ii}$\lambda$4686. This line has been observed in several WR
galaxies, especially in the low-metallicity ones, like Mrk~178 \citep[e.g.,][]
{S99}. A possible explanation for this is that in such
metal-poor environments, WR stars possess sufficiently weak winds that
are optically thin in the He$^{+}$ continuum \citep[e.g.,][]{schmutz92,CH06}, therefore allowing the
He$^{+}$-ionizing photons to escape from the stellar atmospheres.

However, WR features are not always seen when nebular He{\sc ii} is
observed. The nebular emission of He{\sc ii}$\lambda$4686 detected in this
work is shown to be spatially extended, and some He{\sc ii}-emitting
regions are not coincident with the location of the WR bumps by tens
of parsecs (see the map of He{\sc ii}$\lambda$4686 in
Fig.~\ref{maps}). Similar results have been found in
previous work. In \cite{K08} a spatial separation
between the WR star thereabouts and the He{\sc ii} emission is found to be $\sim$ 80
pc in the HII galaxy IIZw70. More recently, \cite{SB12} studying a sample of star-forming galaxies with nebular He{\sc ii}
emission from SDSS DR7 find that a large fraction of the objects at low metallicities do
not show WR signatures. \cite{K11} also identify two He{\sc ii}
nebulae in M33 which do not appear to be associated with any WR star.

Different ionizing mechanisms other than WR stars (e.g. shocks; X-ray
binaries) for the formation of the He{\sc ii} line have been discussed
in the literature \citep[e.g.,][]{gar91,SB12}. Shocks do not seem to be an important excitation source
  in Mrk~178  at large since this galaxy is not detected in X-rays
\citep{stevens98}, indicating that shock-ionization should
not be responsible for the He{\sc ii} emission. Our measured values of
[S{\sc ii}]$\lambda$$\lambda$6717,6731/H$\alpha$ ratio ($\lesssim$
0.20) are lower than the ones observed in SNRs ([S{\sc
  ii}]$\lambda$$\lambda$6717,6731/H$\alpha$ $\sim$ 0.5 - 1.0; \citealt{smith93}] which points against shock ionization as well. The lack of
X-ray detection also leads us to consider X-ray binaries not to be the
main source of He$^{+}$ ionization.

\begin{figure}
\center
\includegraphics[width=0.45\textwidth,clip]{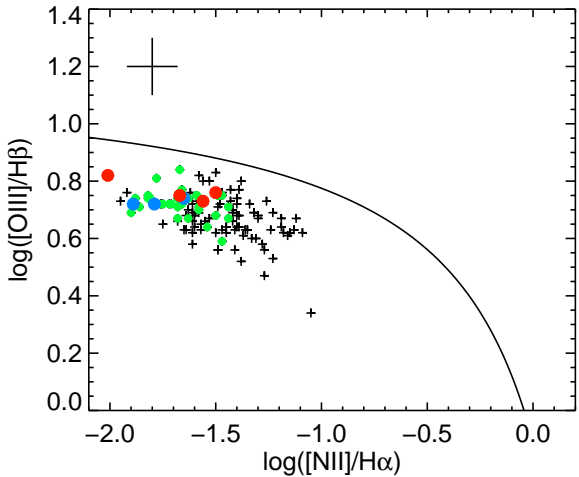}
\caption{log ([O{\sc iii}]$\lambda$5007/H$\beta$) vs. log ([N{\sc ii}]$\lambda$6584/H$\alpha$). Green circles show nebular
He{\sc ii}-emitting spaxels with no WR features, blue circles show the spaxels with blue WR bump
(knot A in Fig.~\ref{maps}) and red circles the spaxels showing both the blue and red WR bumps (knots B and C in Fig.~\ref{maps}). The
remaining spaxels are indicated with cross-symbols. Typical uncertainties on the line ratios are indicated by the cross on the
top-left of the panel. The plotted measurements are corrected for reddening.  The black solid curve is the theoretical
maximum starburst model from Kewley et al. (2001), devised to isolate
objects whose emission line ratios can be accounted for by the photoionization by massive stars (below and to the left of the curve)
from those where some other source of ionization is required.}
\label{bpt} 
\end{figure}

Can we still associate the nebular He{\sc ii} in Mrk~178 with its WR
stars despite the spatial offset between the location of the WR stars
and the He{\sc ii}-emitting zones ? To answer this question we
calculate the gravitational binding energy of a cloud with radius of
$\sim$ 80-100 pc ($\sim$ the maximum observed distances between the WR
stars and nebular He{\sc ii} across the FOV sampled here; see the map
of He{\sc ii}$\lambda$4686 in Fig.~\ref{maps}) and assuming a hydrogen
density of $\sim$ 100 cm$^{-3}$. From this calculation we obtain that
the energy required to excavate a hole of this size is $\sim$
10$^{51-52}$ erg. We checked that this number is of the order of the
predicted mechanical energy injected into the ISM due to massive star
winds taking into account the number of WR stars derived in
section~\ref{wr_stars}. These predictions are based on evolutionary
synthesis models by \cite{M09} for an HII region with Salpeter IMF and
SMC-metallicity. Also, it is worthwhile remembering that WR stars with
high temperatures (T$_{\star}$ $\gtrsim$ 40 kK) and weak winds are
expected to be capable of producing He{\sc ii} ionizing photons
\citep[e.g.,][]{CH06}. WC stars in knots B and C within Mrk~178
certainly appear to satisfy both criteria, although not exactly where
nebular He{\sc ii}4686 is seen, as discussed above. We should bear in
mind that the precise geometry of the ionized ISM distribution  in the close
environment of these knots clearly plays a role.
 If WR stars from
knot A were responsible for the nebular He{\sc ii} emission in this
knot, some of the WN stars would need to be early-types, which cannot
be ruled out from our WR template fits analysis. Additionally,
according to the spectral classiﬁcation scheme indicated in
Fig.~\ref{bpt}, for most positions in Mrk~178 our emission line ratios
fall in the general locus of star-forming objects suggesting that hot
massive stars are the dominant ionizing source within our FOV.

\subsection{Aperture effects}

\cite{BKD08} presented a study of WR galaxies based on
spectra from SDSS DR6 and found Mrk~178 to have the strongest WR
features relative to H$\beta$ of any galaxy in the sample (see their
Figs.18 and 19) which is intriguing if one considers the low
metallicity of Mrk~178. This curious behaviour was possibly attributed
to the fact that, due to the proximity of Mrk~178 (one of the closest
systems in the sample of \citealt{BKD08}), the SDSS fiber could
have by chance hit one single/few luminous WR stars while the H$\beta$
flux is more extended. Actually looking at the H$\alpha$ map in Fig.~\ref{maps} we
see that the Balmer line emission is more extended and its peak is
offset from the WR emission which also indicates propagation of star
formation; where the WR stars are now, the O stars are no longer.
Here we take advantage of the IFS to investigate how the
observations of WR signatures are affected by aperture effects.

\begin{figure}
\center
\includegraphics[width=0.45\textwidth,clip]{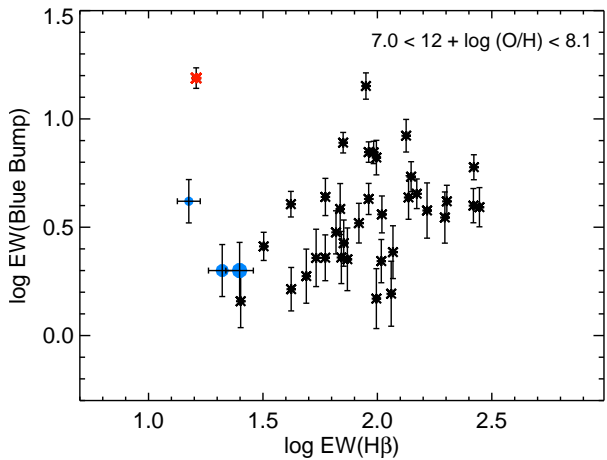}
\caption{EW(blue WR bump) vs EW(H$\beta$). Asterisks show values
  obtained from SDSS DR7 for metal-poor WR galaxies [7.0 $\lesssim$
  12+log(O/H) $\lesssim$ 8.1]; the red asterisk represents Mrk~178. The three blue circles, from the smallest to the biggest one,
  represent the 5, 7 and 10 arcsec-diameter  apertures from
our data centered at knot B.  For the majority of the points the
uncertainty of log[EW(H$\beta$)] is of the order of the symbol size.}
\label{apertures} 
\end{figure}

In Fig.~\ref{apertures} we show EW(WR blue bump) vs EW(H$\beta$)
obtained from SDSS DR7 spectra of low metallicity [7.0 $\lesssim$
12+log(O/H) $\lesssim$ 8.1] WR galaxies and we find the same
discrepancy reported by \cite{BKD08}, i.e. the most deviant
point (red asterisk) in Fig.~\ref{apertures} belongs to the galaxy
Mrk~178. Using our IFU data we construct one-dimensional spectra by
combining spaxels within circular apertures of increasing diameter
(the center of all apertures coincides with our WR Knot B which is
approximately the center of the Mrk~178
SDSS fiber). The measurements derived from these
spectra are blue circles in Fig.~\ref{apertures} which show
that Mrk~178 is brought closer to the bulk of metal-poor systems when
the aperture size increases confirming that the offset between
Mrk~178 and the rest of the objects as viewed by SDSS is not real and caused by
aperture effect instead. We also checked that
for apertures with diameter larger than $\sim$ 10 arcsec we no longer
detect the WR bump. This likely happens because the underlying stellar
absorption masks the WR stellar signatures more easily if the
apertures are too large. In metal-poor systems there is an
enhancement of this effect since WR stars are expected to have weaker
lines than in more metal rich environments \citep{Conti89,CH06}. In
addition, after correcting the EW(WR blue bump) by the physical area
of the SDSS aperture for each
object in Fig.~\ref{apertures}, we verified that the measurements of
EW(WR blue bump) for all galaxies (not
only Mrk~178) suffer from some degree of aperture effects \citep[see
also][]{EPM07}. All this can explain why sometimes WR features
are not detected when they are expected to be present \citep[see
also][]{james10}.

\section{Summary and Conclusions}\label{fim} 

We have presented here the first optical IFS study of the metal-poor
WR galaxy Mrk~178. The proximity of Mrk~178 combined with the IFS
technique allow us to locate and resolve
star-forming knots hosting a few WR stars, and also to characterize the
WR content. In addition we are able to probe the spatial correlation
between massive stars and the properties of the surrounding ISM.  In
the following we list the main results derived from this work.

\begin{itemize}

\item We defined three WR knots from which two are identified for the
  first time here. The WR knot spectra reveal the presence of
  nitrogen-type and carbon-type WR
  stars in Mrk~178. By comparing the observed spectra of the WR knots
  with SMC and LMC template WR stars we estimate a lower limit for the number
  of WR stars ($\gtrsim$ 20) in our Mrk~178 FOV that is already higher
  than that
  currently found in the literature ($\sim$2-3 WR stars from
  \citealt{G00}). This is probably because IFS
not only allows us to sample a larger area of the galaxy, but also offers a more
powerful technique in searching for WR stars.
As was demonstrated in \cite{K08},
  using IFS one can find WR stars where they were not detected before
  because IFS, in comparison with long-slit spectroscopy, can increase the
  contrast of the WR bump emission against the galaxy continuum, thus
  minimizing the WR bump dilution.\\

\item Our statistical analysis suggests that spatial variations in the
gaseous electron temperature exist and that the scatter in T$_{e}$ can
be larger than that in O/H within the observed FOV. Thus, caution
should be exercised 
when analysing integrated spectra of giant HII regions/HII galaxies
which do not necessarily represent the ``local'' ISM properties around
massive star clusters. \\

\item The nebular chemical abundance in Mrk~178 is homogeneous over
  spatial scales of  hundreds of parsecs.
 This result is in agreement with
  previous work on ionized gas in HII galaxies and favours the scheme,
  discussed by \cite{T96}, in which the most recent products
  of massive star nucleosynthesis remain unobservable optically on
  timescales longer than the lifetime of HII regions. The
  representative metallicity of Mrk~178 derived here is
  12+log(O/H)=7.72 $\pm$ 0.01 (error-weighted mean value of O/H and its
  corresponding statistical error). \\

\item To probe the presence of small-scale ($\sim$ 20 pc, our resolution element size) localized chemical
    variations, we performed a close inspection of the chemical
    abundances for the WR knots from which we find a possible
    localized N and He enrichment, spatially correlated with WR knot C.\\

\item The nebular He{\sc ii} emission appears extended and is likely 
  related to ionization from hot stellar continua. We interpret the
  spatial offset between WR stars and nebular He{\sc ii}-emitting regions as
  an effect of the mechanical energy injected by WR star winds This
  agrees with results by \cite{SB12} and \cite{K08}. Caution
    should be exercised when the precise geometry of the ionized ISM
    is not considered. Shock-ionization and X-ray binaries are unlikely to be
  significant ionizing mechanisms since Mrk~178 is not detected in
  X-rays. \\

\item \cite{BKD08} report an unusual (too high) L(WR bump)/L(H$\beta$)
  ratio for Mrk~178 derived from its SDSS spectrum. The authors
  question this assuming that aperture effect can be an
  explanation. This is confirmed in this work with IFS. We also
  demonstrate that using too large an aperture, the chance of 
  detecting WR features decreases. This result
  indicates that WR galaxy samples/catalogues constructed on single
  fiber/long-slit spectrum basis may be biased in
  the sense that WR signatures can escape
  detection depending on the distance of the object  and on the aperture size.
  
\end{itemize}

This paper shows the need to identify good targets, such as Mrk~178, and
the power of IFS for investigating issues related with aperture
effects. In particular, we show the effect of distance and metallicity
on the search for WR features, and the extent to which physical variations in the
ISM properties can be detected. In addition, we believe that this work
makes way for future research of low metallicity, close-by objects
aiming to survey and characterize the WR population and to identify
He{\sc ii} nebulae through high-spatial resolution imaging and IFS.

\section*{Acknowledgements}

We are very grateful to our referee Prof. Anthony Moffat for his fruitful
comments and suggestions on the manuscript.
This work has been partially funded by research projects
AYA2007-67965-C03-02 and AYA2010-21887-C04-01 from the Spanish PNAYA
and CSD2006-00070 1st Science with GTC of the MICINN. RGB acknowledges
support from the Spanish Ministerio de Ciencia e Innovacion through
grant AYA2010-15081. FD acknowledges financial support from CNES.

The STARLIGHT project is supported by the Brazilian agencies CNPq,
CAPES, and FAPESP. This paper uses the plotting package jmaplot
developed by Jes\'us Ma\'{\i}z-Apell\'aniz (available at
http://dae45.iaa.csic.es:8080$\sim$jmaiz/software). This research 
made use of the NASA/IPAC Extragalactic Database (NED) which is
operated by the Jet Propulsion Laboratory, California Institute of
Technology, under contract with the National Aeronautics and Space
Administration.

This paper makes use of the Sloan Digital Sky Survey data.
Funding for SDSS and SDSS-II has been provided by the Alfred P. Sloan
Foundation, the Participating Institutions, the National Science
Foundation, the US Department of Energy, the National Aeronautics and
Space Administration, the Japanese Monbukagakusho, the Max Planck
Society, and the Higher Education Funding Council for England. The
SDSS web site is http://www.sdss.org/. SDSS is managed by the Astrophysical Research Consortium for the
Participating Institutions. The Participating Institutions are the
American Museum of Natural History, Astrophysical Institute Potsdam,
University of Basel, University of Cambridge, Case Western Reserve
University, University of Chicago, Drexel University, Fermilab, the
Institute for Advanced Study, the Japan Participation Group, Johns
Hopkins University, the Joint Institute for Nuclear Astrophysics, the
Kavli Institute for Particle Astrophysics and Cosmology, the Korean
Scientist Group, the Chinese Academy of Sciences (LAMOST), Los Alamos
National Laboratory, the Max-Planck-Institute for Astronomy (MPIA),
the Max-Planck-Institute for Astrophysics (MPA), New Mexico State
University, Ohio State University, University of Pittsburgh,
University of Portsmouth, Princeton University, the United States
Naval Observatory, and the University of Washington.

\bibliographystyle{mn2e}

\begin{thebibliography}{}

\bibitem[\protect\citeauthoryear{Arribas et al.}{1998}]{arribas98} Arribas S., et al., 1998, ASPC, 152, 149

\bibitem[\protect\citeauthoryear{Asplund et al.}{2009}]{asplund09} Asplund M., Grevesse N., Sauval A.~J., Scott P., 2009, ARA\&A, 47, 481

\bibitem[\protect\citeauthoryear{Bingham et al.}{1994}]{bingham94} Bingham R.~G., Gellatly D.~W., Jenkins C.~R., Worswick S.~P., 1994, SPIE, 2198, 56

\bibitem[\protect\citeauthoryear{Brinchmann, Kunth, \& Durret}{2008}]{BKD08} Brinchmann J., Kunth D., Durret F., 2008, A\&A, 485, 657 

\bibitem[\protect\citeauthoryear{{Bruzual} \& {Charlot}}{{Bruzual} \& {Charlot}}{2003}]{bruzual03}{Bruzual} G.,  {Charlot} S.,  2003, \mnras, 344, 1000

\bibitem[\protect\citeauthoryear{Buckalew, Kobulnicky, \& Dufour}{2005}]{B05} Buckalew B.~A., Kobulnicky H.~A., Dufour R.~J., 2005, ApJS, 157, 30 

\bibitem[\protect\citeauthoryear{Cair{\'o}s et al.}{2009a}]{C09mrk409} Cair{\'o}s L.~M., Caon N., Papaderos P., Kehrig C., Weilbacher P., Roth M.~M., Zurita C., 2009a, ApJ, 707, 1676 

\bibitem[\protect\citeauthoryear{Cair{\'o}s et al.}{2009b}]{C09mrk1418} Cair{\'o}s L.~M., Caon N., Zurita C., Kehrig C., Weilbacher P., Roth M., 2009b, A\&A, 507, 1291 

\bibitem[\protect\citeauthoryear{Cair{\'o}s et al.}{2010}]{C10} Cair{\'o}s L.~M., Caon N., Zurita C., Kehrig C., Roth M., Weilbacher P., 2010, A\&A, 520, A90

\bibitem[\protect\citeauthoryear{Campos-Aguilar, Moles, \& Masegosa}{1993}]{campos93} Campos-Aguilar A., Moles M., Masegosa J., 1993, AJ, 106, 1784

\bibitem[\protect\citeauthoryear{{Cardelli}, {Clayton} \& {Mathis}}{{Cardelli} et~al.}{1989}]{cardelli89} {Cardelli} J.~A.,  {Clayton} G.~C.,    {Mathis} J.~S.,  1989, \apj, 345, 245

\bibitem[\protect\citeauthoryear{{Chabrier}}{{Chabrier}}{2003}]{chabrier03}{Chabrier} G.,  2003, \pasp, 115, 763

\bibitem[\protect\citeauthoryear{{Cid Fernandes}, {Gu}, {Melnick}, {Terlevich}, {Terlevich}, {Kunth}, {Rodrigues Lacerda} \& {Joguet}}{{Cid Fernandes} et~al.}{2004}]{cid04}{Cid Fernandes} R.,  {Gu} Q.,  {Melnick} J.,  {Terlevich} E.,  {Terlevich} R., {Kunth} D.,  {Rodrigues Lacerda} R., {Joguet} B.,  2004, \mnras, 355, 273

\bibitem[\protect\citeauthoryear{{Cid Fernandes}, {Mateus}, {Sodr{\'e}}, {Stasi{\'n}ska} \& {Gomes}}{{Cid Fernandes} et~al.}{2005}]{cid05}{Cid Fernandes} R.,  {Mateus} A.,  {Sodr{\'e}} L.,  {Stasi{\'n}ska} G., {Gomes} J.~M.,  2005, \mnras, 358, 363

\bibitem[\protect\citeauthoryear{Conti, Garmany, \& Massey}{1989}]{Conti89} Conti P.~S., Garmany C.~D., Massey P., 1989, ApJ, 341, 113 

\bibitem[\protect\citeauthoryear{Crowther et al.}{1995}]{C95} Crowther P.~A., Smith L.~J., Hillier D.~J., Schmutz W., 1995, A\&A, 293, 427

\bibitem[\protect\citeauthoryear{Crowther, De Marco, \& Barlow}{1998}]{C98} Crowther P.~A., De Marco O., Barlow M.~J., 1998, MNRAS, 296, 367

\bibitem[\protect\citeauthoryear{Crowther et al.}{2003}]{Crowther03} Crowther, P.~A., Drissen, L., Abbott, J.~B.,  Royer, P., Smartt, S.~J.  2003, A\&A 404, 483

\bibitem[\protect\citeauthoryear{Crowther \& Hadfield}{2006}]{CH06} Crowther P.~A., Hadfield L.~J., 2006, A\&A, 449, 711 

\bibitem[\protect\citeauthoryear{Crowther et al.}{2007}]{C07} Crowther P.~A., Carpano S., Hadfield L.~J., Pollock A.~M.~T., 2007, A\&A, 469, L31

\bibitem[\protect\citeauthoryear{Crowther}{2007}]{PC07} Crowther P.~A., 2007, ARA\&A, 45, 177 

\bibitem[\protect\citeauthoryear{de Mello et al.}{1998}]{D98} de Mello D.~F., Schaerer D., Heldmann J., Leitherer C., 1998, ApJ, 507, 199 

\bibitem[\protect\citeauthoryear{De Vaucouleurs et al.}{1991}]{rc3} DeVaucouleurs G., De Vaucouleurs A., Corwin H.~G., Jr Buta R.~ J.,Paturel G., Fouque P., 1991, Third Reference Catalogue of Bright Galaxies, Version 3.9

\bibitem[\protect\citeauthoryear{Drissen et al.}{1995}]{drissen95} Drissen L., Moffat A.~F.~J., Walborn, N.~R., Shara M.~M., 1995, AJ, 110, 2235


\bibitem[\protect\citeauthoryear{Eldridge \& Stanway}{2009}]{eldridge09} Eldridge J.~J., Stanway E.~R., 2009, MNRAS, 400, 1019 

\bibitem[\protect\citeauthoryear{Esteban \& Vilchez}{1992}]{EV92} Esteban C., Vilchez J.~M., 1992, ApJ, 390, 536 

\bibitem[\protect\citeauthoryear{Fern{\'a}ndez-Mart{\'{\i}}n et al.}{2012}]{alba12} Fern{\'a}ndez-Mart{\'{\i}}n A., Mart{\'{\i}}n-Gord{\'o}n D., V{\'{\i}}lchez J.~M., P{\'e}rez Montero E.,Riera A., S{\'a}nchez S.~F., 2012, A\&A, 541, A119 

\bibitem[\protect\citeauthoryear{Foellmi, Moffat, \& Guerrero}{2003}]{fmg} Foellmi C., Moffat A.~F.~J., Guerrero M.~A., 2003, MNRAS, 338, 360

\bibitem[\protect\citeauthoryear{Foellmi, Moffat, \& Guerrero}{2003}]{foellmi03} Foellmi C., Moffat A.~F.~J., Guerrero M.~A., 2003, MNRAS, 338, 1025 

\bibitem[\protect\citeauthoryear{Garc{\'{\i}}a-Benito et al.}{2010}]{RGB10} Garc{\'{\i}}a-Benito R., et al., 2010, MNRAS, 408, 2234 

\bibitem[\protect\citeauthoryear{Garc{\'{\i}}a-Benito \& P{\'e}rez-Montero}{2012}]{RGB12} Garc{\'{\i}}a-Benito R., P{\'e}rez-Montero E., 2012, MNRAS, 423, 406 

\bibitem[\protect\citeauthoryear{Garnett et al.}{1991}]{gar91} Garnett D.~R., Kennicutt R.~C., Jr., Chu Y.-H., Skillman E.~D., 1991, ApJ, 373, 458 

\bibitem[\protect\citeauthoryear{{Gil de Paz}, {Madore} \& {Pevunova}}{{Gil de  Paz} et~al.}{2003}]{gildepaz03}
{Gil de Paz} A.,  {Madore} B.~F., {Pevunova} O.,  2003, \apjs, 147, 29

\bibitem[\protect\citeauthoryear{Gil de Paz \& Madore}{2005}]{gildepaz05} Gil de Paz A., Madore B.~F., 2005, ApJS, 156, 345

\bibitem[\protect\citeauthoryear{Gonzalez-Delgado et al.}{1994}]{rosa94} Gonzalez-Delgado R.~M., et al., 1994, ApJ, 
437, 239

\bibitem[\protect\citeauthoryear{Gonzalez-Delgado et al.}{1996}]{rosa96} Gonzalez-Delgado R.~M., et al., 1996, ApJ, 473, 1125 

\bibitem[\protect\citeauthoryear{Gonzalez-Riestra, Rego, \& Zamorano}{1988}]{G88} Gonzalez-Riestra R., Rego M., Zamorano J., 1988, A\&A, 202, 27 

\bibitem[\protect\citeauthoryear{Guseva, Izotov, \& Thuan}{2000}]{G00} Guseva N.~G., Izotov Y.~I., Thuan T.~X., 2000, ApJ, 531, 776

\bibitem[\protect\citeauthoryear{Hadfield \& Crowther}{2006}]{HC06} Hadfield L.~J., Crowther P.~A., 2006, MNRAS, 368, 1822 

\bibitem[\protect\citeauthoryear{Izotov, Thuan, \& Lipovetsky}{1994}]{I94} Izotov Y.~I., Thuan T.~X., Lipovetsky V.~A., 1994, ApJ, 435, 647

\bibitem[\protect\citeauthoryear{Izotov, Thuan, \& Stasi{\'n}ska}{2007}]{I07} Izotov Y.~I., Thuan T.~X., Stasi{\'n}ska G., 2007, ApJ, 662, 15 

\bibitem[\protect\citeauthoryear{Izotov et al.}{2009}]{I09} Izotov Y.~I., Guseva N.~G., Fricke K.~J., Papaderos P., 2009, A\&A, 503, 61

\bibitem[\protect\citeauthoryear{James, Tsamis, \& Barlow}{2010}]{james10} James B.~L., Tsamis Y.~G., Barlow M.~J., 2010, MNRAS, 401, 759

\bibitem[\protect\citeauthoryear{Karachentsev et al.}{2003}]{kar03} Karachentsev I.~D., et al., 2003, A\&A, 398, 467

\bibitem[\protect\citeauthoryear{Kehrig, Telles, \& Cuisinier}{2004}]{K04} Kehrig C., Telles E., Cuisinier F., 2004, AJ, 128, 1141

\bibitem[\protect\citeauthoryear{Kehrig et al.}{2006}]{K06} Kehrig C., V{\'{\i}}lchez J.~M., Telles E., Cuisinier F., P{\'e}rez-Montero E., 2006, A\&A, 457, 477 

\bibitem[\protect\citeauthoryear{Kehrig et al.}{2008}]{K08} Kehrig C., V{\'{\i}}lchez J.~M., S{\'a}nchez S.~F., Telles E., P{\'e}rez-Montero E., Mart{\'{\i}}n-Gord{\'o}n D., 2008, A\&A, 477, 813 

\bibitem[\protect\citeauthoryear{Kehrig et al.}{2011}]{K11} Kehrig C., et al., 2011, A\&A, 526, A128 

\bibitem[\protect\citeauthoryear{Kewley et al.}{2001}]{kew01} Kewley L.~J., Dopita M.~A., Sutherland R.~S., Heisler C.~A., Trevena J., 2001, ApJ, 556, 121 

\bibitem[\protect\citeauthoryear{Kingsburgh, Barlow, \& Storey}{1995}]{kingsburgh95} Kingsburgh R.~L., Barlow M.~J., Storey P.~J., 1995, A\&A, 295, 75

\bibitem[\protect\citeauthoryear{Kingsburgh \& Barlow}{1995}]{kingsburghdr1} Kingsburgh R.~L., Barlow M.~J., 1995, A\&A, 295, 171 

\bibitem[\protect\citeauthoryear{Kobulnicky \& Skillman}{1996}]{kob96} Kobulnicky H.~A., Skillman E.~D., 1996, ApJ, 471, 211 

\bibitem[\protect\citeauthoryear{Kobulnicky \& Skillman}{1997}]{KS97} Kobulnicky H.~A., Skillman E.~D., 1997, ApJ, 489, 636

\bibitem[\protect\citeauthoryear{Kobulnicky et al.}{1997}]{kob97} Kobulnicky H.~A., Skillman E.~D., Roy J.-R., Walsh J.~R., Rosa M.~R., 1997, ApJ, 477, 679 

\bibitem[\protect\citeauthoryear{Kunth \& Sargent}{1981}]{KS81} Kunth D., Sargent W.~L.~W., 1981, A\&A, 101, L5

\bibitem[\protect\citeauthoryear{{Le Borgne}, {Bruzual}, {Pell{\'o}}, {Lan{\c c}on}, {Rocca-Volmerange}, {Sanahuja}, {Schaerer}, {Soubiran} \& {V{\'{\i}}lchez-G{\'o}mez}}{{Le Borgne} et~al.}{2003}]{leborgne03}
{Le Borgne} J.-F.,  {Bruzual} G.,  {Pell{\'o}} R.,  {Lan{\c c}on} A., {Rocca-Volmerange} B.,  {Sanahuja} B.,  {Schaerer} D.,  {Soubiran} C.,{V{\'{\i}}lchez-G{\'o}mez} R.,  2003, \aap, 402, 433

\bibitem[\protect\citeauthoryear{Lebouteiller et al.}{2009}]{lebouteiller09} Lebouteiller V., Kunth D., Thuan T.~X., D{\'e}sert J.~M., 2009, A\&A, 494, 915

\bibitem[\protect\citeauthoryear{Lee \& Skillman}{2004}]{lee04} Lee H., Skillman E.~D., 2004, ApJ, 614, 698

\bibitem[\protect\citeauthoryear{Lee et al.}{2009}]{lee09} Lee J.~C., et al., 2009, ApJ, 706, 599 

\bibitem[\protect\citeauthoryear{Legrand et al.}{1997}]{L97} Legrand F., Kunth D., Roy J.-R., Mas-Hesse J.~M., Walsh J.~R., 1997, A\&A, 326, L17 

\bibitem[\protect\citeauthoryear{{Leitherer}, {Schaerer}, {Goldader},{Delgado}, {Robert}, {Kune}, {de Mello}, {Devost} \& {Heckman}}{{Leitherer}et~al.}{1999}]{leitherer99}{Leitherer} C.,  {Schaerer} D.,  {Goldader} J.~D.,  {Delgado} R.~M.~G., {Robert} C.,  {Kune} D.~F.,  {de Mello} D.~F.,  {Devost} D., {Heckman} T.~M.,  1999, \apjs, 123, 3

\bibitem[\protect\citeauthoryear{Lilliefors}{1967}]{li67} Lilliefors, H.~W., 1967, J. Amer. Statistical Assoc., 62, 399

\bibitem[\protect\citeauthoryear{L{\'o}pez-S{\'a}nchez et al.}{2011}]{lopez11} L{\'o}pez-S{\'a}nchez {\'A}.~R., Mesa-Delgado A., L{\'o}pez-Mart{\'{\i}}n L., Esteban C., 2011, MNRAS,411, 2076

\bibitem[\protect\citeauthoryear{Maeder}{1992}]{maeder92} Maeder A., 1992, A\&A, 264, 105

\bibitem[\protect\citeauthoryear{Markarian}{1969}]{mrk69} Markarian B.~E., 1969, Afz, 5, 581 

\bibitem[\protect\citeauthoryear{{Mateus}, {Sodr{\'e}}, {Cid Fernandes}, {Stasi{\'n}ska}, {Schoenell} \& {Gomes}}{{Mateus} et~al.}{2006}]{mateus06} {Mateus} A.,  {Sodr{\'e}} L.,  {Cid Fernandes} R.,  {Stasi{\'n}ska} G.,{Schoenell} W.,    {Gomes} J.~M.,  2006, \mnras, 370, 721

\bibitem[\protect\citeauthoryear{Melnick, Terlevich, \& Eggleton}{1985}]{M85} Melnick J., Terlevich R., Eggleton P.~P., 1985, MNRAS, 216, 255 

\bibitem[\protect\citeauthoryear{Meynet \& Maeder}{2005}]{mm05} Meynet G., Maeder A., 2005, A\&A, 429, 581

\bibitem[\protect\citeauthoryear{Modjaz et al.}{2008}]{modjaz08} Modjaz M., et al., 2008, AJ, 135, 1136

\bibitem[\protect\citeauthoryear{Moffat, Seggewiss \& Shara}{1985}]{Moffat85} Moffat, A~F.~J., Seggewiss W., Shara, M.~M. 1985, ApJ 295, 109

\bibitem[\protect\citeauthoryear{Moll et al.}{2007}]{moll07} Moll S.~L., Mengel S., de Grijs R., Smith L.~J., Crowther P.~A., 2007, MNRAS, 382, 1877

\bibitem[\protect\citeauthoryear{Moll{\'a}, Garc{\'{\i}}a-Vargas, \& Bressan}{2009}]{M09} Moll{\'a} M., Garc{\'{\i}}a-Vargas M.~L., Bressan A., 2009, MNRAS, 398, 451 

\bibitem[\protect\citeauthoryear{Monreal-Ibero, Walsh, \& V{\'{\i}}lchez}{2012}]{ana12} Monreal-Ibero A., Walsh J.~R., V{\'{\i}}lchez J.~M., 2012, A\&A, 544, A60

\bibitem[\protect\citeauthoryear{Munoz-Tunon et al.}{1996}]{cassiana96} Munoz-Tunon C., Tenorio-Tagle G., Castaneda H.~O., Terlevich R., 1996, AJ, 112, 1636 

\bibitem[\protect\citeauthoryear{Neugent \& Massey}{2011}]{NM11} Neugent K.~F., Massey P., 2011, ApJ, 733, 123 

\bibitem[\protect\citeauthoryear{Olive \& Skillman}{2004}]{olive04} Olive K.~A., Skillman E.~D., 2004, ApJ, 617, 29

\bibitem[\protect\citeauthoryear{Pagel, Terlevich, \& Melnick}{1986}]{pagel86} Pagel B.~E.~J., Terlevich R.~J., Melnick J., 1986, PASP, 98, 1005 

\bibitem[\protect\citeauthoryear{Papaderos \& {\"O}stlin}{2012}]{P12} Papaderos P., {\"O}stlin G., 2012, A\&A, 537, A126 

\bibitem[\protect\citeauthoryear{{P{\'e}rez-Montero} \& {D{\'{\i}}az}}{{P{\'e}rez-Montero} \& {D{\'{\i}}az}}{2003}]{EPM03}{P{\'e}rez-Montero} E.,  {D{\'{\i}}az} A.~I.,  2003, \mnras, 346, 105

\bibitem[\protect\citeauthoryear{P{\'e}rez-Montero \& D{\'{\i}}az}{2005}]{EPM05} P{\'e}rez-Montero E., D{\'{\i}}az A.~I., 2005, MNRAS, 361, 1063 

\bibitem[\protect\citeauthoryear{P{\'e}rez-Montero \& D{\'{\i}}az}{2007}]{EPM07} P{\'e}rez-Montero E., D{\'{\i}}az {\'A}.~I., 2007, MNRAS, 377, 1195 

\bibitem[\protect\citeauthoryear{P{\'e}rez-Montero \& Contini}{2009}]{EPMC09} P{\'e}rez-Montero E., Contini T., 2009, MNRAS, 398, 949 

\bibitem[\protect\citeauthoryear{P{\'e}rez-Montero et al.}{2009}]{EPM09} P{\'e}rez-Montero E., Garc{\'{\i}}a-Benito R., D{\'{\i}}az A.~I., P{\'e}rez E., Kehrig C., 2009, A\&A, 497, 53 

\bibitem[\protect\citeauthoryear{P{\'e}rez-Montero et al.}{2010}]{EPM10} P{\'e}rez-Montero E., Garc{\'{\i}}a-Benito R., H{\"a}gele G.~F., D{\'{\i}}az {\'A}.~I., 2010, MNRAS, 404, 2037 

\bibitem[\protect\citeauthoryear{P{\'e}rez-Montero et al.}{2011}]{EPM11} P{\'e}rez-Montero E., et al., 2011, A\&A, 532, A141 

\bibitem[\protect\citeauthoryear{Popescu \& Hopp}{2000}]{popescu00} Popescu C.~C., Hopp U., 2000, A\&AS, 142, 247 

\bibitem[\protect\citeauthoryear{Recchi et al.}{2003}]{recchi03} Recchi S., Matteucci F., D'Ercole A., Tosi M., 2003, Ap\&SS, 284, 623 

\bibitem[\protect\citeauthoryear{Recchi \& Hensler}{2013}]{recchi13} Recchi S., Hensler G., 2013, arXiv, arXiv:1301.0812

\bibitem[\protect\citeauthoryear{Roy \& Kunth}{1995}]{RK95} Roy J.-R., Kunth D., 1995, A\&A, 294, 432

\bibitem[\protect\citeauthoryear{Russell \& Dopita}{1990}]{RD90} Russell S.~C., Dopita M.~A., 1990, ApJS, 74, 93

\bibitem[\protect\citeauthoryear{Sandin et al.}{2010}]{sandin10} Sandin C., Becker T., Roth M.~M., Gerssen J., Monreal-Ibero A., B{\"o}hm P., Weilbacher P., 2010, A\&A, 515, A35

\bibitem[\protect\citeauthoryear{Schaerer, Contini, \& Pindao}{1999}]{S99} Schaerer D., Contini T., Pindao M., 1999, A\&AS, 136, 35

\bibitem[\protect\citeauthoryear{{Schlegel}, {Finkbeiner} \& {Davis}}{{Schlegel} et~al.}{1998}]{schlegel98} {Schlegel} D.~J.,  {Finkbeiner} D.~P.,    {Davis} M.,  1998, \apj, 500, 525

\bibitem[\protect\citeauthoryear{Schmutz, Leitherer, \& Gruenwald}{1992}]{schmutz92} Schmutz W., Leitherer C., Gruenwald R., 1992, PASP, 104, 1164 

\bibitem[\protect\citeauthoryear{Schnurr et al.}{2008}]{oli} Schnurr O., Moffat A.~F.~J., St-Louis N., Morrell N.~I., Guerrero M.~A., 2008, MNRAS, 389, 806 

\bibitem[\protect\citeauthoryear{Shaw \& Dufour}{1994}]{shaw94} Shaw R.~A., Dufour R.~J., 1994, ASPC, 61, 327

\bibitem[\protect\citeauthoryear{Shirazi \& Brinchmann}{2012}]{SB12} Shirazi M., Brinchmann J., 2012, MNRAS, 421, 1043 

\bibitem[\protect\citeauthoryear{Sidoli, Smith, \& Crowther}{2006}]{sidoli96} Sidoli F., Smith L.~J., Crowther P.~A., 2006, MNRAS, 370, 799

\bibitem[\protect\citeauthoryear{Smith}{1973}]{smith73} Smith L.~F., 1973, IAUS, 49, 15 

\bibitem[\protect\citeauthoryear{Smith et al.}{1993}]{smith93} Smith R.~C., Kirshner R.~P., Blair W.~P., Long K.~S., Winkler P.~F., 1993, ApJ, 407, 564

\bibitem[\protect\citeauthoryear{Smith, Shara, \& Moffat}{1996}]{smith96} Smith L.~F., Shara M.~M., Moffat A.~F.~J., 1996, MNRAS, 281, 163 

\bibitem[\protect\citeauthoryear{Stevens \& Strickland}{1998}]{stevens98} Stevens I.~R., Strickland D.~K., 1998, MNRAS, 294, 523 

\bibitem[\protect\citeauthoryear{Stock \& Barlow}{2010}]{stockbarlow10} Stock D.~J., Barlow M.~J., 2010, MNRAS, 409, 1429 

\bibitem[\protect\citeauthoryear{{Storey} \& {Hummer}}{{Storey} \& {Hummer}}{1995}]{storey95}{Storey} P.~J.,  {Hummer} D.~G.,  1995, \mnras, 272, 41

\bibitem[\protect\citeauthoryear{Tenorio-Tagle}{1996}]{T96} Tenorio-Tagle G., 1996, AJ, 111, 1641

\bibitem[\protect\citeauthoryear{van Zee  \& Haynes}{2006}]{vanZ06} van Zee L., Haynes M.~P., 2006, ApJ, 636, 214 (HII regions in dwarf irregular galaxies)

\bibitem[\protect\citeauthoryear{V{\'{\i}}lchez \& Iglesias-P{\'a}ramo}{1998}]{V98} V{\'{\i}}lchez J.~M., Iglesias-P{\'a}ramo J., 1998, ApJ, 508, 248 

\bibitem[\protect\citeauthoryear{York et al.}{2000}]{y00} York D.~G., et al., 2000, AJ, 120, 1579

\bibitem[\protect\citeauthoryear{Westera et al.}{2004}]{W04} Westera P., Cuisinier F., Telles E., Kehrig C., 2004, A\&A, 423, 133

\bibitem[\protect\citeauthoryear{Westera et al.}{2012}]{W12} Westera P., Cuisinier F., Curty D., Buser R., 2012, MNRAS, 421, 398

\bibitem[\protect\citeauthoryear{Woosley \& Bloom}{2006}]{WB06} Woosley S.~E., Bloom J.~S., 2006, ARA\&A, 44, 507

\end{thebibliography}

\label{lastpage}
\end{document}